\documentclass[12pt]{article}
\usepackage{a4p}
\usepackage{epsfig}
\usepackage{amstex}     % math niceties

\begin{document}
%=======================================================================
%       Parameters for the title page
%=======================================================================
%
%  PPE number, Date and Version
%
\newcommand{\PPEnum}    {CERN-EP/98-107}
\newcommand{\Date}      {3rd July 1998}
\newcommand{\Author}    {Shoji Asai and Sachio Komamiya}
\newcommand{\MailAddr}  {Shoji.Asai@@cern.ch}
\newcommand{\EdBoard}   {Benno List, Homer Neal, Hartmut Rick and \\
Claire Shepherd-Themistocleous}
\newcommand{\DraftVer}  {Version 5.1}
\newcommand{\DraftDate} {02 July 1998}
\newcommand{\TimeLimit} {2 July 1998; 17:00 h}
\newcommand{\Submit} {To be submitted to Euro. Phys. J. C}

% Modified POINT
%   
 
%=======================================================================
%       Parameters affecting the general appearance
%=======================================================================
\def\toprule{\noalign{\hrule \medskip}}
\def\midrule{\noalign{\medskip\hrule }}
\def\botrule{\noalign{\medskip\hrule }}
\setlength{\parskip}{\medskipamount}
 
%=======================================================================
%       Define symbols
%=======================================================================

\newcommand{\ee}{{\mathrm e}^+ {\mathrm e}^-}
\newcommand{\sq}{\tilde{\mathrm q}}
\newcommand{\seff}{\tilde{\mathrm f}}
\newcommand{\sele}{\tilde{\mathrm e}}
\newcommand{\sell}{\tilde{\ell}}
\newcommand{\snu}{\tilde{\nu}}
\newcommand{\ch}{\tilde{\chi}^\pm}
\newcommand{\chp}{\tilde{\chi}_{1}^+}
\newcommand{\chm}{\tilde{\chi}_{1}^-}
\newcommand{\chpm}{\tilde{\chi}_{1}^\pm}
\newcommand{\nt}{\tilde{\chi}^0}
\newcommand{\qq}{{\mathrm q}\bar{\mathrm q}}
\newcommand{\qqx}{{\mathrm q}\bar{\mathrm q'}}
\newcommand{\nunu}{\nu \bar{\nu}}
\newcommand{\mumu}{\mu^+ \mu^-}
\newcommand{\tautau}{\tau^+ \tau^-}
\newcommand{\ellell}{\ell^+ \ell^-}
\newcommand {\wenu} {{\mathrm{We}} \nu}
\newcommand{\nulqq}{\nu \ell {\mathrm q} \bar{\mathrm q}'}
\newcommand{\MZ}{M_{\mathrm Z}}
\newcommand{\MW}{M_{\mathrm W}}

\newcommand {\stopm}         {\tilde{\mathrm{t}}_{1}}
\newcommand {\stops}         {\tilde{\mathrm{t}}_{2}}
\newcommand {\stopbar}       {\bar{\tilde{\mathrm{t}}}_{1}}
\newcommand {\stopx}         {\tilde{\mathrm{t}}}
\newcommand {\sneutrino}     {\tilde{\nu}}
\newcommand {\slepton}       {\tilde{\ell}}
\newcommand {\stopl}         {\tilde{\mathrm{t}}_{\mathrm L}}
\newcommand {\stopr}         {\tilde{\mathrm{t}}_{\mathrm R}}
\newcommand {\stoppair}      {\tilde{\mathrm{t}}_{1} \bar{\tilde{\mathrm{t}}}_{1}}
\newcommand {\gluino}        {\tilde{\mathrm g}}

\newcommand {\sbotm}         {\tilde{\mathrm{b}}_{1}}
\newcommand {\sbots}         {\tilde{\mathrm{b}}_{2}}
\newcommand {\sbotbar}       {\bar{\tilde{\mathrm{b}}}_{1}}
\newcommand {\sbotx}         {\tilde{\mathrm{b}}}
\newcommand {\sbotl}         {\tilde{\mathrm{b}}_{\mathrm L}}
\newcommand {\sbotr}         {\tilde{\mathrm{b}}_{\mathrm R}}
\newcommand {\sbotpair}      {\tilde{\mathrm{b}}_{1} \bar{\tilde{\mathrm{b}}}_{1}}

\newcommand {\neutralino}    {\tilde{\chi }^{0}_{1}}
\newcommand {\neutrala}      {\tilde{\chi }^{0}_{2}}
\newcommand {\neutralb}      {\tilde{\chi }^{0}_{3}}
\newcommand {\neutralc}      {\tilde{\chi }^{0}_{4}}
\newcommand {\bino}          {\tilde{\mathrm B}^{0}}
\newcommand {\wino}          {\tilde{\mathrm W}^{0}}
\newcommand {\higginoa}      {\tilde{\mathrm H_{1}}^{0}}
\newcommand {\higginob}      {\tilde{\mathrm H_{1}}^{0}}
\newcommand {\chargino}      {\tilde{\chi }^{\pm}_{1}}
\newcommand {\charginop}     {\tilde{\chi }^{+}_{1}}
\newcommand {\KK}            {{\mathrm K}^{0}-\bar{\mathrm K}^{0}}
\newcommand {\ff}            {{\mathrm f} \bar{\mathrm f}}
\newcommand {\bq}            {\mathrm b} 
\newcommand {\cq}            {\mathrm c} 
\newcommand {\ele}           {\mathrm e} 
\newcommand {\bstopm} {\mbox{$\boldmath {\tilde{\mathrm{t}}_{1}} $}}
\newcommand {\Mt}            {M_{\mathrm t}}
\newcommand {\mb}            {M_{\mathrm b}}
\newcommand {\mc}            {M_{\mathrm c}}
\newcommand {\mscalar}       {m_{0}}
\newcommand {\Mgaugino}      {M_{1/2}}
\newcommand {\tanb}          {\tan \beta}
\newcommand {\rs}            {\sqrt{s}}
\newcommand {\WW}            {{\mathrm W}^+{\mathrm W}^-}
\newcommand {\eetautau}      {\ee-\rightarrow {\tau^+}{\tau^-}}
\newcommand {\MGUT}          {M_{\mathrm{GUT}}}
\newcommand {\Zboson}        {{\mathrm Z}^{0}}
\newcommand {\Wpm}           {{\mathrm W}^{\pm}}
\newcommand {\Wp}            {{\mathrm W}^{+}}
\newcommand {\allqq}         {\sum_{q \neq t} q \bar{q}}
\newcommand {\mixstop}       {\theta _{\stopx}}
\newcommand {\mixsbot}       {\theta _{\sbotx}}
\newcommand {\phiacop}       {\phi _{\mathrm {acop}}}
\newcommand {\cosjet}        {\cos\thejet}
\newcommand {\costhr}        {\cos\thethr}
\newcommand {\djoin}         {d_{\mathrm{join}}}
\newcommand {\mchar}         {m_{\chpm}}
\newcommand {\mstop}         {m_{\stopm}}
\newcommand {\msbot}         {m_{\sbotm}}
\newcommand {\mchi}          {m_{\neutralino}}
\newcommand {\pp}{p \bar{p}}
 
% Def. fuer groesser-ungefaehr:
\newcommand{\gsim}{\;\raisebox{-0.9ex}
           {$\textstyle\stackrel{\textstyle >}{\sim}$}\;}
% Def. fuer kleiner-ungefaehr:
\newcommand{\lsim}{\;\raisebox{-0.9ex}{$\textstyle\stackrel{\textstyle<}
           {\sim}$}\;}

\newcommand{\degree}    {^\circ}
%
%-----------------------------------------
%  Variables for machine; math mode only
%-----------------------------------------
\newcommand{\Ecm}       {E_{\mathrm{cm}}}
\newcommand{\Ebeam}     {E_{\mathrm{b}}}
\newcommand{\roots}     {\sqrt{s}}
%----------------------------------------
%  Variables for events; math mode only
%----------------------------------------
%
%     Thrust
%
\newcommand{\thrust}    {T}
\newcommand{\nthrust}   {\hat{n}_{\mathrm{thrust}}}
\newcommand{\thethr}    {\theta_{\,\mathrm{thrust}}}
\newcommand{\phithr}    {\phi_{\mathrm{thrust}}}
\newcommand{\acosthr}   {|\cos\thethr|}
\newcommand{\thejet}    {\theta_{\,\mathrm{jet}}}
\newcommand{\acosjet}   {|\cos\thejet|}
\newcommand{\thmiss}    { \theta_{\mathrm{miss}} }
\newcommand{\cosmiss}   {| \cos \thmiss |}
\newcommand{\pbinv}     {\mathrm{pb}^{-1}}
%
%     Energy, etc.
%
\newcommand{\Evis}      {E_{\mathrm{vis}}}
\newcommand{\Rvis}      {E_{\mathrm{vis}}\,/\roots}
\newcommand{\Mvis}      {M_{\mathrm{vis}}}
\newcommand{\Rbal}      {R_{\mathrm{bal}}}
\newcommand{\mjet}      {\bar{M}_{\mathrm{jet}}}
%
%---------
%  Units
%---------
%----------------------------
%  Bibliographic references
%----------------------------
%
%     Journal names
%
\newcommand{\PhysLett}  {Phys.~Lett.}
\newcommand{\PRL} {Phys.~Rev.\ Lett.}
\newcommand{\PhysRep}   {Phys.~Rep.}
\newcommand{\PhysRev}   {Phys.~Rev.}
\newcommand{\NPhys}  {Nucl.~Phys.}
\newcommand{\NIM} {Nucl.~Instr.\ Meth.}
\newcommand{\CPC} {Comp.~Phys.\ Comm.}
\newcommand{\ZPhys}  {Z.~Phys.}
\newcommand{\IEEENS} {IEEE Trans.\ Nucl.~Sci.}
%
%     Collaboration names
%
\newcommand{\OPALColl}  {OPAL Collab.}
\newcommand{\JADEColl}  {JADE Collab.}
\newcommand{\etal}      {{\it et~al.}}
%-------
%  etc
%-------
\newcommand{\onecol}[2] {\multicolumn{1}{#1}{#2}}
\newcommand{\ra}        {\rightarrow}   % \to can be used as well
 
% Def. fuer groesser-ungefaehr:
%\newcommand{\gsim}{\;\raisebox{-0.9ex}
%           {$\textstyle\stackrel{\textstyle >}{\sim}$}\;}
% Def. fuer kleiner-ungefaehr:
%\newcommand{\lsim}{\;\raisebox{-0.9ex}{$\textstyle\stackrel{\textstyle<}
%           {\sim}$}\;}
 
%=======================================================================
%       Title Page
%=======================================================================
 
%-----------------------------------------------------------------------
 
\begin{titlepage}
%
%     Header
%
\begin{center}
    \large
    EUROPEAN LABORATORY FOR PARTICLE PHYSICS
\end{center}
\begin{flushright}
    \large
    \PPEnum\\
%    \PNnum\\
    \Date
\end{flushright}
%
%       Draft notice
%
%% \ifthenelse{\boolean{Draft}} {
%%    \begin{center}
%%      \Large\sf\bf
%%      Draft \DraftVer,\ \ \DraftDate \\
%%    \end{center}
%%    \vspace*{3mm}
%% } {}
%
%     Main title
%
\begin{center}
    \huge\bf\boldmath
    Search for Scalar Top \\
    and Scalar Bottom Quarks \\
    at $\rs$ = 183~GeV at LEP
\end{center}
\bigskip
\bigskip
\bigskip
%
%     Author names
%
\begin{center}
    \LARGE
    The OPAL Collaboration \\
%\bigskip 
%\bigskip
%\large
% {Authors: \Author \\
% Editorial board: \EdBoard}
\end{center}
\bigskip
\bigskip
\bigskip
%
%     Abstract
%
\begin{abstract}%=======================================================
Searches for a scalar top quark and a scalar bottom quark 
have been performed using a total data sample of 56.8~pb$^{-1}$
at a centre-of-mass energy of $\roots = $183~GeV
collected with the OPAL detector at LEP.
No candidate events were observed.
Combining this result with those obtained at lower 
centre-of-mass energies,
the 95\% C.L. lower limit on the scalar top quark mass is  
85.0~GeV if the mixing angle between the supersymmetric partners 
of the left- and right-handed states of the top quark is zero.
The lower limit is 81.3~GeV,
even if the scalar top quark decouples from the $\Zboson$ boson. 
These limits were obtained assuming that
the scalar top quark decays into 
a charm quark and the lightest neutralino,
and that the mass difference between 
the scalar top quark and the lightest 
neutralino is larger than 10~GeV.
%Assuming a relatively light scalar neutrino, 
The complementary decay mode of the scalar top quark 
in which it decays into a bottom quark, 
a charged lepton and a scalar neutrino was also studied. 
From a similar analysis, a mass limit on the light scalar bottom quark 
was set at 82.7~GeV for
a mass difference between the scalar bottom quark and the lightest 
neutralino larger than 7~GeV,
and at 84.0~GeV for the mass difference larger 10~GeV and 
the lightest neutralino heavier than 30~GeV\@. 
These limits were obtained assuming that
the scalar bottom quark decays into 
a bottom quark and the lightest neutralino,
and that a mixing angle between the supersymmetric partners 
of the left- and right-handed states of the bottom quark is zero.
\end{abstract}%=========================================================
 
\bigskip
\bigskip
\begin{center}
{\large (\Submit) }\\
%\bigskip
%\bigskip
%{\large \bf Draft \DraftVer, \DraftDate}
\end{center}
 
%\smallskip
%\begin{center}
%{\large \bf Comments to \MailAddr}\\
%{\large \bf by \TimeLimit}\\
%\end{center}
 
%\smallskip
%\begin{center}
%{\large \bf This note describes preliminary OPAL results,
%and is intended for the members of the collaboration}\\
%\end{center}
 
\end{titlepage}
\begin{center}{\Large        The OPAL Collaboration
}\end{center}\bigskip
\begin{center}{
%begin authorlist
K.\thinspace Ackerstaff$^{  8}$,
G.\thinspace Alexander$^{ 23}$,
J.\thinspace Allison$^{ 16}$,
N.\thinspace Altekamp$^{  5}$,
K.J.\thinspace Anderson$^{  9}$,
S.\thinspace Anderson$^{ 12}$,
S.\thinspace Arcelli$^{  2}$,
S.\thinspace Asai$^{ 24}$,
D.\thinspace Axen$^{ 29}$,
G.\thinspace Azuelos$^{ 18,  a}$
% ...
%end authorlist
}\end{center}\bigskip
\bigskip
%begin institutes
$^{  1}$School of Physics and Space Research, University of Birmingham,
Birmingham B15 2TT, UK
\newline
$^{  2}$Dipartimento di Fisica dell' Universit\`a di Bologna and INFN,
I-40126 Bologna, Italy
\newline
% ...
$^{  5}$Cavendish Laboratory, Cambridge CB3 0HE, UK
\newline
% ...
$^{  8}$CERN, European Organisation for Particle Physics,
CH-1211 Geneva 23, Switzerland
\newline
$^{  9}$Enrico Fermi Institute and Department of Physics,
University of Chicago, Chicago IL 60637, USA
\newline
% ...
$^{ 12}$Indiana University, Department of Physics,
Swain Hall West 117, Bloomington IN 47405, USA
\newline
% ...
$^{ 16}$Department of Physics, Schuster Laboratory, The University,
Manchester M13 9PL, UK
\newline
% ...
$^{ 18}$Laboratoire de Physique Nucl\'eaire, Universit\'e de Montr\'eal,
Montr\'eal, Quebec H3C 3J7, Canada
\newline
% ...
$^{ 23}$Department of Physics and Astronomy, Tel Aviv University,
Tel Aviv 69978, Israel
\newline
$^{ 24}$International Centre for Elementary Particle Physics and
Department of Physics, University of Tokyo, Tokyo 113, and
Kobe University, Kobe 657, Japan
\newline
% ...
$^{ 29}$University of British Columbia, Department of Physics,
Vancouver BC V6T 1Z1, Canada
\newline
% ...
$^{  a}$ and at TRIUMF, Vancouver, Canada V6T 2A3
\newline
% ...
\newline
 
%=======================================================================
\section{Introduction}
%=======================================================================

Supersymmetric (SUSY) extensions \cite{SUSY}  of the Standard Model
predict the existence of the bosonic partners of all known fermions.
The scalar top quark~($\stopx$), which is the bosonic partner of the 
top quark, can be the lightest charged supersymmetric 
particle for the following two reasons~\cite{stop1}\@.
Firstly, one-loop radiative corrections to the
$\stopx$ mass through Higgsino-quark loops and
Higgs-squark loops are always negative.
The correction is large for a heavy top quark mass as 
measured by the CDF and D0 Collaborations~\cite{CDF}\@.
Secondly, the supersymmetric partners of the
right-handed and left-handed top quarks
($\stopr$ and $\stopl$) mix, and
the resultant two mass eigenstates ($\stopm$ and $\stops$)
have a mass splitting.
This mass splitting is expected to be very large 
due to the large top quark mass.  
It is possible that the lighter mass eigenstate 
($\stopm$) can be lighter than any other
charged SUSY particle, and also lighter than the top quark itself.
The $\stopm$ is the mixed state of $\stopr$ and $\stopl$,
{\it i.e.} $\stopm = \stopl \cos \mixstop + \stopr \sin \mixstop$,
where $\mixstop$ is a mixing angle.
All SUSY breaking parameters \cite{SUSY} are hidden 
in the $\mixstop$ and a mass of $\stopm$\@.

The scalar bottom quark ($\sbotx$) 
can also be light if $\tanb$, the ratio of 
vacuum expectation values of the two Higgs doublets, is larger than 
approximately 40.
In this case, analogous mixing between the supersymmetric partners 
of the right- and left-handed states 
of the bottom quark ($\sbotr$ and $\sbotl$) becomes large,
and the resultant two mass eigenstates ($\sbotm$ and $\sbots$)
also have a large mass splitting~\cite{bartl}\@.
The mass of the lighter mass eigenstate 
($\sbotm$) may therefore be within the reach of LEP2. 

Scalar top quark pairs and scalar bottom quark pairs 
are produced in $\ee$ annihilation
via a virtual $\Zboson$ boson or a virtual photon.
In this paper it is assumed that
R-parity~\cite{RP} is conserved and that either  
$\neutralino$ or $\snu$ is the only SUSY particle which is
lighter than  $\stopm$ ($\sbotm$),
where $\neutralino$ is 
the lightest neutralino and $\snu$ is the scalar neutrino.
The dominant decay mode of the $\stopm$ with the above assumptions is  
expected to be either\footnote{Through out this paper,
all references to particle or decay implicitly include charge conjugation.}
$\stopm \ra \cq \neutralino$ 
or $\stopm \ra \bq \snu \ell^{+} $\@.
Both of these decay modes have been searched for. 
The dominant decay mode of the $\sbotm$ is 
expected to be $ \sbotm \ra \bq \neutralino$\@.
Under the assumption of  R-parity conservation,
the $\neutralino$ and $\snu$ are invisible in the detector.
Thus, $\stoppair$ and $\sbotpair$ events 
are characterised by two acoplanar 
jets\footnote{Two 
jets not back-to-back with each other in the plane 
perpendicular to the beam axis.} 
or two acoplanar jets plus
two leptons, with missing energy.

The D0 Collaboration has reported a lower limit~\cite{d0} 
on the $\stopm$ mass of about 85~GeV (95\% C.L.)
for the case of $\stopm \ra \cq \neutralino$
and that the mass difference 
between $\stopm$ and $\neutralino$ is larger than about 35~GeV\@.
Searches at $\ee$ colliders are sensitive to a smaller mass difference 
and mass limits for the $\stopm$ have been obtained around 
the $\Zboson$ peak (LEP1) assuming $\stopm \ra \cq \neutralino$\@.
A 95\% C.L. lower limit of about 45~GeV was obtained for a 
mass difference larger than 5~GeV~\cite{opalstop}.
Previous searches at centre-of-mass energies 
of $\rs$= 130, 136 ~\cite{stop133},
161~\cite{stop161} and 
171~GeV ~\cite{stop171,alephstop} improved the limit 
on the mass of the $\stopm$ to 66.8~GeV 
if the $\mixstop$ is 
smaller than $\pi/4$ and if the mass difference 
between $\stopm$ and $\neutralino$ is larger 
than 10~GeV.

In 1997 the LEP $\ee$ collider at CERN was run 
at centre-of-mass energies of 181--184~GeV\@.
The luminosity weighted mean centre-of-mass energy was 182.7~GeV\@.
In this paper direct searches for $\stopm$ and $\sbotm$
using the data collected with the OPAL detector at these centre-of-mass 
energies are reported.
The results shown here have been obtained by combining the results
obtained at these new centre-of-mass energies 
with those obtained at $\rs$ 
= 130, 136, 161 and 171~GeV \cite{stop161,stop171}\@. 

The phenomenology of the production and decay of 
the $\stopm$ ($\sbotm$) is described in section 2 of 
the previous publication~\cite{stop171}\@.
In this paper, 
the OPAL detector and the event simulation   
for signal and background processes are given in section 2.
In section 3, the data analysis is described and the results are given in
section 4. 

%=======================================================================
\section{The OPAL Detector and Event Simulation}
%=======================================================================
 
%=======================================================================
\subsection{The OPAL Detector}
%=======================================================================

The OPAL detector,
which is described in detail in ref.~\cite{OPAL-detector},
is a multipurpose apparatus
having nearly complete solid angle coverage.
The central detector consists of
a silicon strip detector and tracking chambers,
providing charged particle tracking 
for over 96\% of the full solid 
angle,
inside a uniform solenoidal magnetic field of 0.435~T\@.
A lead-glass electromagnetic (EM) calorimeter  
located outside the magnet coil
is hermetic
in the polar angle\footnote{A right-handed 
coordinate system is adopted,
where the $x$-axis points to the centre of the LEP ring,
and positive $z$ is along  the electron beam direction.
The angles $\theta$ and $\phi$ are the polar and azimuthal angles,
respectively.} range of $|\cos \theta |<0.82$ for the barrel
region and $0.81<|\cos \theta |<0.984$ for the endcap region.
The magnet return yoke
consisting of barrel and endcap sections along with
pole tips 
is instrumented for hadron calorimetry (HCAL)
in the region $|\cos \theta |<0.99$\@.
Four layers of muon chambers cover the outside of the hadron calorimeter.
Calorimeters close to the beam axis measure the luminosity
using small angle Bhabha scattering events
and complete the geometrical acceptance down to 24 mrad.
These include 
the forward detectors (FD) which are
lead-scintillator sandwich calorimeters and, at smaller angles,
silicon tungsten calorimeters (SW)~\cite{SW}
located on both sides of the interaction 
point.
Tungsten shields were installed around the beam pipe
in front of the SW detectors to reduce the amount of synchrotron radiation
seen by the detector. The presence of the shield results in a gap
in the SW acceptance between the polar angles of 28 and 31~mrad.
The gap between the endcap EM calorimeter and the FD
is filled by an additional lead-scintillator 
electromagnetic calorimeter,
called the gamma-catcher.

\subsection{Monte Carlo Event Simulation of $\stopm$ and $\sbotm$}

Monte Carlo simulation of the production and decay of the $\stopm$
was performed as follows~\cite{stopgen}\@.
The $\stoppair$ pairs were generated
taking into account initial-state radiation~\cite{lund}\@.
The hadronisation process was subsequently performed
to produce colourless $\stopm$-hadrons and other fragmentation products
according to the Lund string fragmentation scheme
(JETSET 7.4)~\cite{lund,fragment}\@.
The parameters for perturbative QCD and fragmentation processes
were optimised using event shape
distributions of the hadronic $\Zboson$ decays
measured by OPAL~\cite{opalfragment}\@.
For the fragmentation of the $\stopm$,
the fragmentation function proposed
by Peterson {\it et al.}~\cite{lund,Peterson} was used,
where the parameter $\epsilon_{\stopm}$ was set to 
\begin{equation}
 \epsilon_{\stopm}=\epsilon_{\bq}
m_{\bq}^{2} / {\mstop}^{\!\!\!2}  \ \ \ 
( \epsilon_{\bq} \ = \ 0.0038 ~\cite{opalfragment}, \ 
m_{\bq} \ = \ 5 \ {\mathrm {GeV}} ) \  .
\end{equation}

The $\stopm$-hadron was formed from a $\stopm$-quark 
and a spectator anti-quark or diquark \cite{spectator}\@.
The fragmentation products excluding the $\stopm$-hadrons
carry less than 2\% of the centre-of-mass energy. 
For the $\stopm$ decaying into $\cq \neutralino$,
a colour string was stretched between the charm quark and
the spectator.
This colour singlet system 
was hadronised using the Lund scheme~\cite{lund,fragment}\@.
Gluon bremsstrahlung (QCD parton showering) was allowed
in this process, and 
the Peterson function was also used 
for the charm quark fragmentation,
where $\epsilon_{\cq} $ was set to 0.031~\cite{opalfragment}\@. 
The signals for the decays $ \stopm \ra \bq \ell^{+} \snu $ 
and $\sbotm \ra \bq \neutralino$ 
were simulated in a similar manner.

One thousand events were generated at each of
56 combinations of $(m_{\stopm}, m_{\neutralino})$ for
$ \stopm \ra \cq \neutralino$, 
40 combinations of  $(m_{\stopm}, m_{\snu})$ for 
$\stopm \ra \bq \ell^{+} \snu$
and 40 combinations of  $(m_{\stopm}, m_{\snu})$ for 
$\stopm \ra \bq \tau^{+} \snu$\@.
The $\sbotpair$ events were generated for 48 combinations
of $(m_{\sbotm}, m_{\neutralino})$\@.
The mixing angles of the $\stopm$ and $\sbotm$ were set to
zero when these events were generated.
The dependence of the detection efficiencies 
on these mixing angles will be discussed in section~4.1\@.
The generated events were processed
through the full simulation of the OPAL detector~\cite{GOPAL},
and the same event analysis chain was applied to the simulated events
and the data.

\subsection{Monte Carlo Event Simulation of Background Processes}
 
The background processes were simulated as follows:

\noindent $\bullet$ Multijet hadronic events $\ee \ra \qq (\gamma) $ 
in which one or two jet momenta are mismeasured are 
a background for the large $\Delta m$ region 
($\equiv \, \mstop - \mchi , \, \mstop - m_{\snu}$, or $\msbot - \mchi$)\@.
The PYTHIA~\cite{lund} generator was used to simulate hadronic events.
 
\noindent $\bullet$ $\tau$ pairs, in which one of the
$\tau$ lepton decays into a low momentum electron and energetic neutrinos,
are a background to acoplanar two-jet events.
The KORALZ event generator~\cite{KORALZ} was used for the
generation of $\tau^+ \tau^- (\gamma)$ and $\mumu (\gamma)$ events.  
The BHWIDE program~\cite{BHWIDE} was used for $\ee \ra \ee (\gamma) $ 
events. 
  
\noindent $\bullet$ Two-photon processes are the most important background
for the case of a small mass difference $\Delta m $, 
since such signal events have small visible energy and 
small transverse momentum relative to
the beam direction.
Using the Monte Carlo generators 
PHOJET~\cite{PHOJET}, PYTHIA~\cite{lund} and HERWIG~\cite{HERWIG},
hadronic events from two-photon processes were simulated
in which the invariant mass of the photon-photon
system ($M_{\gamma \gamma}$) was larger than 5.0~GeV\@.
Monte Carlo samples for four-lepton events ($\ee \ee$, $\ee \mumu$ and 
$\ee \tautau$) were generated with the Vermaseren 
program~\cite{Vermaseren}\@.  
 
\noindent $\bullet$ Finally, four-fermion processes in which at least one
of the fermions is a neutrino constitute a serious background.
The dominant contributions come from  
${\mathrm W}^+ {\mathrm W}^-$ or $\gamma ^{*} \Zboson $ events.
Since the interference effects of various diagrams are important,
the grc4f generator~\cite{grace} was used,  
which takes into account all interfering diagrams and 
includes initial-state photon radiation.

These background events were also processed
through the full detector simulation
and the same event analysis chain as used for the data 
was applied.
 
%=======================================================================
\section{Analysis}
%=======================================================================
 
The present analysis is based on the data
collected during the 1997 run of LEP2\@.
% The data used in this analysis correspond
% to an integrated luminosity of 56.8~pb$^{-1}$
% at centre-of-mass energies of $\roots = $181--184~GeV\@.
% The luminosity weighted mean centre-of-mass energy is 182.7~GeV\@.
Since the event topologies of $\stopm \ra \cq \neutralino$ 
and $\sbotm \ra \bq \neutralino $ are similar, 
the same selection criteria were used for both these modes 
(section~3.1  analysis A)\@. 
In section 3.2 (analysis B), the selection criteria for
$\stopm \ra \bq \ell^{+} \snu$ are discussed. 
These analyses are similar to those in ref.~\cite{stop171},
and the same quality criteria as in ref.~\cite{stop171} were used
to select tracks and clusters.

Variables used for the cuts, such as 
the total visible energy, $\Evis$, the total transverse momentum
and the acoplanarity angle (defined below) were calculated as follows.
First, the four-momenta of the tracks and 
those of the EM and HCAL clusters not
associated with charged tracks were summed.
Whenever a calorimeter cluster had associated charged tracks,
the expected energy deposited by the tracks was subtracted
from the cluster energy to reduce double counting.
If the energy of a cluster was smaller
than the expected energy deposited by the associated tracks,
the cluster energy was not used.
Hadron calorimeter clusters were also used in calculating 
event variables.
A large momentum-unbalance is occasionally caused by
the fluctuation in the energy measurement of clusters 
in the hadron calorimeter
because of the limited energy resolution.
Therefore the transverse momentum and the visible mass 
calculated without the HCAL clusters were also used
to reduce two-photon background processes.

The following preselection criteria~(P1 -- P2),
which are common to both analyses A and B,
were applied first. 
The numbers of events remaining after each cut are listed
in Table~\ref{tab:nevA}\@.
For comparison, the table also shows 
the corresponding numbers of simulated events
for background processes and 
for three samples of the simulated $\stoppair$ 
($\stopm \ra \cq \neutralino$) and $\sbotpair$ events.
 
\begin{description}

\item[(P1)]
The number of charged tracks was required to be at least four. 
The ratio of the number of good tracks
to the total number of reconstructed tracks was required 
to be greater than 0.2 
to reduce beam-gas and beam-wall background events.
The visible mass of the event, excluding the hadron calorimeter,
was also required to be larger than 3~GeV\@.
 
\item[(P2)]
To reduce the background from two-photon processes and multihadronic
events, where a jet axis was close to the beam direction, the total energy
deposited had to be less than 5~GeV in each SW detector,
less than 2~GeV in each FD detector 
and less than 5~GeV in each side of the gamma-catcher.

\end{description}

\subsection{Analysis A: {\boldmath$\boldsymbol{\stopm \ra \cq \neutralino}$} 
and {\boldmath$\boldsymbol{\sbotm \ra \bq \neutralino}$}} 
 
The experimental signature for 
$\stoppair$($\stopm \ra \cq \neutralino$) events
and $\sbotpair$ events
is an acoplanar two-jet topology with a large transverse 
momentum with respect to the beam axis.
The fragmentation functions of $\stopm $ and $\sbotm $ are very 
hard and the invariant mass of charm (bottom) quark and
the spectator quark is small,
therefore the jets are expected to be narrow.
 
The event selection criteria are described below.
The numbers of events remaining after each cut are also listed
in Table~\ref{tab:nevA}\@.

\begin{description}

\item[(A1)]
The visible energy in the region of $|\cos \theta|>0.9$
was required to be less than 20\% of the total visible energy.
In addition,
the polar angle of the missing momentum direction, $\thmiss$,
was required to satisfy $\cosmiss < 0.9$
to reduce the two-photon and beam-gas events.
 
\item[(A2)]
Events from two-photon processes were largely removed
by demanding that
the event transverse momentum excluding the hadron calorimeter, 
$P_t$, be greater than 4.5~GeV and that the transverse momentum
including the hadron calorimeter, $P_t^{\mathrm{HCAL}}$,
be greater than 4.5~GeV\@.
Fig.~1 shows the distribution of $P_t$ just before these cuts.

\item[(A3)]
The number of reconstructed jets was required to be exactly two.
Jets were reconstructed using the Durham algorithm~\cite{DURHAM} 
with the jet resolution parameter of 
$y_{\rm cut}$ = $0.005 (\Evis / \rs)^{-1}$\@.
This $\Evis$-dependent $y_{\rm cut}$ parameter
was necessary for good jet-reconstruction over a wide range of 
$\mstop$, $\msbot$ and $\mchi$\@.
Fig.~2 shows the number of reconstructed jets before this cut. 
Furthermore, both reconstructed jets were required to contain 
at least two charged particles
to reduce the $\tau^{+} \tau^{-}$ background where
at least one of the $\tau$'s decayed into only one charged particle.

\item[(A4)]
The acoplanarity angle, $\phiacop$,
is defined as $\pi$ minus the azimuthal opening angle
between the directions of the two reconstructed jets.
To ensure the reliability of the calculation of $\phiacop$,
both jet axes were required to satisfy $|\cos{\theta}_{\rm jet}| < 0.95$,
where ${\theta}_{\rm jet}$ is the polar angle of the jet.
The value of $\phiacop$ was required to be greater than 20$\degree$\@.
Fig.~3 shows the distributions of $\phiacop$ just before
this selection.

\item[(A5)]
`Softness' was defined as 
($\frac{{M}_1}{{E}_1} + \frac{{M}_2}{{E}_2}$),
where ${M}_1$ and ${M}_2$ are the invariant masses 
of the two reconstructed jets, 
and ${E}_1$ and ${E}_2$ are the energies of the jets.
The signal events have low values of `Softness',  
on the other hand, the two-photon events,
which pass through the acoplanarity cut, have relatively large values. 
It was required that $1.5 \times {\rm Softness} < (P_t-4.5)$,
where $P_t$ is given in units of GeV\@.
Fig.~4 shows the scatter plots of Softness versus $P_t$ for 
data, the simulated two-photon events, 
the $\stoppair$ and $\sbotpair$ events.

\item[(A6)]
The arithmetic mean of the invariant masses of the jets, $\mjet$, 
was required to be smaller than 8~GeV\@.
When the invariant mass of the event, $\Mvis$, was larger than 65~GeV,
a harder cut $\mjet < $ 5~GeV was applied to 
reduce background from $\wenu$ events.
Fig.~5 shows the scatter plots of $\Mvis$ versus $\mjet$ for 
data, the simulated four-fermion events,
the $\stoppair$ and $\sbotpair$ events.
As shown in Fig.~5(c) and (d), 
jets from the $\stopm$ and $\sbotm$ are expected to have
low invariant masses, 
because the fragmentation function of the $\stopm$ is hard
and only a few particles are emitted  
from the fragmentation process of $\stoppair$.

\end{description}

\begin{table}[h]%------------------------------------------------------
\centering
\begin{tabular}{|l||r||r||r|r|r|r||r|r|r|}
\hline%-----------------------------------------------------------------
   &  \multicolumn{1}{c||}{data} & \multicolumn{1}{c||}{total} &
$\qq (\gamma)$ & $\ellell (\gamma)$ & 
\multicolumn{1}{|c|}{`$\gamma \gamma$'} &
\multicolumn{1}{c||}{4-f} &
\multicolumn{3}{c|}{\rule{0mm}{6mm} $\stoppair$ and $\sbotpair$}  \\
    &    &   \multicolumn{1}{c||}{bkg.} 
&  &  &  &  & \multicolumn{3}{c|}{ }   \\
\hline%-----------------------------------------------------------------
$m_{\stopm}$ (GeV)&       &      &         &          &            &
        & \multicolumn{1}{c|}{80} & 
          \multicolumn{1}{c|}{80} &
          \multicolumn{1}{c|}{--}  \\
%------------------------------------------------------------------
$m_{\sbotm}$ (GeV)&       &      &         &          &            &
        & \multicolumn{1}{c|}{--} & 
          \multicolumn{1}{c|}{--} &
          \multicolumn{1}{c|}{80}  \\
%------------------------------------------------------------------
$m_{\neutralino}$ (GeV)&       &      &         &            &            &
        & \multicolumn{1}{c|}{75} & 
          \multicolumn{1}{c|}{60} & 
          \multicolumn{1}{c|}{60}  \\
\hline%-----------------------------------------------------------------
 Presel. 1
& 275501 & 250435 &  5028 &  1229 & 243137 & 1042 & 874 & 967 & 994 \\
\hline% ----------------------------------------
 Presel. 2
& 141619 & 124394 &  3618 &  1165 & 118753 &  859 & 848 & 933 & 960 \\
\hline% ----------------------------------------
 cut (A1)
&  22927 &  19583 &  1126 &   236 &  17651 &  569 & 731 & 768 & 809 \\
\hline% ----------------------------------------
 cut (A2)
&  1441 &    1461 &   771 &   156 &   47.1 &  488 & 458 & 745 & 778 \\
\hline% ----------------------------------------
 cut (A3)
&   362 &     359 &   298 &  10.4 &   16.3 & 34.0 & 305 & 649 & 717 \\
\hline% ----------------------------------------
 cut (A4)
&    15 &    25.3 &  0.09 &  0.06 &  3.41  & 21.8 & 289 & 582 & 638 \\
\hline% ----------------------------------------
 cut (A5)
&    15 &    22.3 &  0.08 &  0.06 &  0.38  & 21.7 & 229 & 582 & 636 \\
\hline% ----------------------------------------
 cut (A6)
&     0 &    1.97 &  0.04 &  0.04 &  0.27  & 1.63 & 229 & 576 & 609 \\
\hline% ----------------------------------------
\end{tabular}
\caption[]
{
The remaining numbers of events
normalised to the
integrated luminosity of the data
for various background processes are compared with data
after each cut.
Numbers for 
three simulated event samples of $\stoppair$ and $\sbotpair$ are 
also given (each starting from 1000 events).
Before the cut (A2) was applied,
there is a discrepancy between data and the simulated background processes,
since the numbers of events expected from two-photon processes
do not include the region
$M_{\gamma \gamma} <5$~GeV\@.
}
\label{tab:nevA}
\end{table}%------------------------------------------------------------
 
\begin{table}[h]%------------------------------------------------------
\centering
\begin{tabular}{|l||r|r|r|r|r|r|r|r|}
\hline%-----------------------------------------------------------------
~~~~~~~$m_{\stopm}$ (GeV) &  50 &  55 &  60 &  65 &  70 &  75 &  80 &  85 \\
\hline%-----------------------------------------------------------------
\hline%-----------------------------------------------------------------
$\Delta m$            &     &     &     &     &     &     &     &     \\
\hline%-----------------------------------------------------------------
 3~GeV              &  11 &   9 &   7 &   6 &   5 &   3 &   2 &   2 \\
\hline%-----------------------------------------------------------------
 5~GeV              &  29 &  30 &  27 &  28 &  27 &  25 &  23 &  22 \\
\hline%-----------------------------------------------------------------
10~GeV              &  40 &  44 &  45 &  48 &  49 &  48 &  48 &  49 \\
\hline%-----------------------------------------------------------------
20~GeV              &  36 &  45 &  49 &  52 &  53 &  58 &  58 &  60 \\
\hline%-----------------------------------------------------------------
 $\mstop$/2           &  35 &  38 &  41 &  45 &  44 &  47 &  51 &  52 \\
\hline%-----------------------------------------------------------------
 $\mstop $ -- 10~GeV  &  27 &  28 &  28 &  29 &  32 &  35 &  40 &  40 \\
\hline%-----------------------------------------------------------------
 $\mstop$             &  24 &  27 &  29 &  29 &  33 &  33 &  37 &  36 \\
\hline%-----------------------------------------------------------------
\end{tabular}
\caption[]{
The detection efficiencies in percent for $\stoppair$,
in which $\stopm$ decays into $\cq \neutralino$ for different
$\stopm$ masses and $\Delta m$ values,
where $\Delta m$ is $m_{\stopm}-m_{\neutralino}$\@.
The statistical fluctuations of these efficiencies are about 2\% (absolute)\@. 
}
\label{tab:effcn}
\end{table}%------------------------------------------------------------

\begin{table}[h]%------------------------------------------------------
\centering
\begin{tabular}{|l||r|r|r|r|r|r|r|r|}
\hline%-----------------------------------------------------------------
~~~~~~~$m_{\sbotm}$ (GeV) &  50 &  55 &  60 &  65 &  70 &  75 &  80 &  85 \\
\hline%-----------------------------------------------------------------
\hline%-----------------------------------------------------------------
$\Delta m$            &     &     &     &     &     &     &     &     \\
\hline%-----------------------------------------------------------------
 7~GeV              &  34 &  34 &  36 &  36 &  35 &  38 &  37 &  36 \\
\hline%-----------------------------------------------------------------
10~GeV              &  40 &  43 &  46 &  46 &  50 &  49 &  50 &  49 \\
\hline%-----------------------------------------------------------------
20~GeV              &  36 &  41 &  48 &  51 &  55 &  59 &  61 &  63 \\
\hline%-----------------------------------------------------------------
 $\msbot$/2           &  29 &  34 &  35 &  37 &  37 &  41 &  47 &  50 \\
\hline%-----------------------------------------------------------------
 $\msbot $ -- 10~GeV  &  22 &  22 &  22 &  25 &  25 &  27 &  32 &  33 \\
\hline%-----------------------------------------------------------------
 $\msbot$             &  21 &  23 &  21 &  22 &  23 &  26 &  31 &  31 \\
\hline%-----------------------------------------------------------------
\end{tabular}
\caption[]{
The detection efficiencies in percent for $\sbotpair$ for different
$\sbotm$ masses and $\Delta m$ values,
where $\Delta m=m_{\sbotm}-m_{\neutralino}$\@.
The statistical fluctuations of these efficiencies are about 2\% (absolute)\@. 
}
\label{tab:effbn}
\end{table}%------------------------------------------------------------

After all the cuts, no events were observed in the data,
which is consistent with the expected number of background events of 2.0\@.
The four-fermion processes are the dominant background processes.
Uncertainties in the number of background for these processes 
will be discussed in Section~4.2\@.

The efficiencies for $\stoppair$ and $\sbotpair $ events are listed 
in Tables~\ref{tab:effcn} and ~\ref{tab:effbn}\@. 
Both efficiencies are 30--60\%,
if the mass difference between the $\stopm$($\sbotm$) 
and $\neutralino $ is larger than 10~GeV\@.
A modest efficiency of about 20\% is also obtained for 
a mass difference of 5~GeV for $\stoppair$ events.
In addition to effects included in the detector simulation, an
additional efficiency loss of 3.6\% (relative) arose from beam-related
background in the SW, FD and gamma-catcher detectors 
estimated using random beam crossing events.
The efficiencies given in
Tables 1--5 do not include this correction, but it is included when
deriving the mass limits.
The efficiency at an arbitrary point was estimated using a polynomial fit to
the efficiencies determined using the Monte Carlo simulations.

\subsection{Analysis B: {\boldmath$\boldsymbol{\stopm \ra \bq \ell \snu}$}} 
 
The experimental signature for $\stoppair$($\stopm \ra \bq \ell \snu$) events
is an two acoplanar jets plus two leptons with missing transverse 
momentum with respect to the beam axis.
The momenta of leptons and the missing transverse momentum
depend strongly on the mass difference
between $\stopm$ and $\snu$\@.
To obtain optimal performance, 
two sets of selection criteria (analyses B-L and B-H) 
were applied depending on this mass difference.
If the mass difference is smaller than or equal to 10~GeV,
the momenta of leptons and the missing transverse momentum
are relatively small. 
In such cases, it is difficult to identify leptons effectively and
the dominant background comes from two-photon processes.
Selections were optimised to reduce these two-photon events.
When the mass difference is larger than 10~GeV,
the momenta of leptons and the missing transverse momentum
become large. 
In such cases the four-fermion processes become the dominant background
processes.

The numbers of events remaining after each cut are listed
in Tables~\ref{tab:nevBL} and \ref{tab:nevBH}\@.
For comparison, the table also shows 
the corresponding numbers for simulated events
for background processes and
for two samples of simulated $\stoppair$ events, in which 
the branching fraction to each lepton flavour is assumed to be the same.

\subsubsection{Small mass difference case} 

For the case of small mass difference ($\Delta m \leq$ 10~GeV), 
the following selections were applied.

\begin{description}

\item[(B-L1)]
The visible energy in the region of $|\cos \theta|>0.9$
was required to be less than 10\% of the total visible energy.
In addition, $\cosmiss < 0.8$ was required.

\item[(B-L2)]
Both $P_t$ and $P_t^{\mathrm{HCAL}}$ were required 
to be greater than 5~GeV\@.

\item[(B-L3)] 
The number of charged tracks was required to be at least six.
Furthermore,
the number of reconstructed jets was required to be at least four,
because the signal should contain two hadronic jets 
plus two isolated leptons.
Jets were reconstructed using the Durham algorithm~\cite{DURHAM} with 
the jet resolution parameter of $y_{\mathrm{cut}}$ = 0.004\@.
Fig.~6 shows the distributions of the number of reconstructed 
jets before this selection.
  
\item[(B-L4)] 
To examine the acoplanarity of the events, jets were 
reconstructed using the Durham algorithm
where the number of jets was forced to be two.
To ensure a good measurement of acoplanarity angle,
$|\cos{\theta}_{\rm jet}| < 0.95$ was required for 
both reconstructed jets.
Finally, the acoplanarity angle between these two jets
was required to be greater than 15$\degree$\@.
In the three-body decay, the transverse momentum carried by the $\snu$ 
with respect to the original $\stopm$-momentum 
is smaller than that of $\neutralino$ in the two-body decay.
When the $\stopm$ is light, the outgoing $\snu$ is strongly boosted 
toward the direction of the parent $\stopm$\@.
Hence $\phiacop$ for the signal becomes small.
This is the reason for the use of a looser acoplanarity angle cut. 

\item[(B-L5)] 
The total visible energy normalised to the centre-of-mass energy, 
$\Evis/\rs$, was required to be smaller than 0.2 to reject
four-fermion events.
  
\end{description}

No events were observed in the data after the above cuts.
The number of expected background events is 1.1\@.
Two-photon processes are the dominant background.
Uncertainties in the number of expected background events
will be discussed in Section~4.2\@.
 
\begin{table}[h]%------------------------------------------------------
\centering
\begin{tabular}{|l||r||r||r|r|r|r||r|r|}
\hline%-----------------------------------------------------------------
   &  \multicolumn{1}{c||}{data}         & \multicolumn{1}{c||}{total} 
   &  \multicolumn{1}{c|}{$\qq (\gamma)$} 
   &  \multicolumn{1}{c|}{$\ellell (\gamma)$} 
   &  \multicolumn{1}{c|}{`$\gamma \gamma$'} 
   &  \multicolumn{1}{c||}{4-f} 
   &  \multicolumn{2}{c|}{\rule{0mm}{6mm} $\stoppair$}  \\
   &       & \multicolumn{1}{c||}{bkg.} 
   &  &  &  &  & \multicolumn{2}{c|}{ }   \\
\hline %-----------------------------------------------------------------
$m_{\stopm}$ (GeV)&       &      &         &          &            &
        & \multicolumn{1}{c|}{80} & 
          \multicolumn{1}{c|}{80} \\
%------------------------------------------------------------------
$m_{\snu}$ (GeV)&       &      &         &            &            &
        & \multicolumn{1}{c|}{73} & 
          \multicolumn{1}{c|}{70} \\
\hline %------------------------------------------------------------
 cut (B-L1)
& 11713 & 9354 &   904 &   134 &  7859 &   457 & 717 & 695 \\
\hline% ----------------------------------------
 cut (B-L2)
&  1064 & 1076 &   592 &  78.3 &  18.2 &   388 & 139 & 440 \\
\hline% ----------------------------------------
 cut (B-L3)
&   287 &  295 &  94.8 &  0.02 &  1.84 &   198 & 137 & 435 \\
\hline% ----------------------------------------
 cut (B-L4)
&    60 & 62.6 &  3.33 &  0.00 &  0.97 &  58.3 & 118 & 379 \\
\hline% ----------------------------------------
 cut (B-L5)
&     0 & 1.07 &  0.01 &  0.00 &  0.97 &  0.09 & 118 & 378 \\
\hline% ----------------------------------------
\end{tabular}
\caption[]{
The remaining numbers of events
normalised to the
integrated luminosity of the data
for various background processes are compared with data
after each cut.
Numbers for
two simulated event samples of $\stoppair$ are 
also given (each starting from 1000 events).
In these samples, the branching fraction to each lepton flavour
is assumed to be the same.
Before the cut (B-L2),
there is a discrepancy between data and the simulated background processes,
since the numbers of events expected from two-photon processes
do not include the region
$M_{\gamma \gamma} <5$~GeV\@.
}
\label{tab:nevBL}
\end{table}%------------------------------------------------------------
 
\subsubsection{Large mass difference case} 

The selections for a large mass difference ($\Delta m > $ 10~GeV), 
are described below.
Because the momenta of the leptons for this case 
are high enough to be identified,
it was required that events contained at least one lepton.
Then the other cuts (B-H1 and B-H3) were relaxed compared to
the small mass difference case.   

\begin{description}

\item[(B-H1)]
Cut (A1) was applied to reduce two-photon and beam-gas events. 

\item[(B-H2)]
Cut (B-L2) was applied to reject two-photon events.

\item[(B-H3)] 
The number of charged tracks was required to be at least six, 
and the number of reconstructed jets was required 
to be at least three.
Jets were reconstructed with 
the jet resolution parameter of $y_{\mathrm{cut}}$ = 0.004\@.
 
\item[(B-H4)] 
Cut (B-L4) was applied to reject multihadronic events. 

\item[(B-H5)] 
A candidate event was required to contain at least one lepton.
Leptons were identified in the following way:
electrons were selected if they satisfied either of the two
identification methods described in Ref.~\cite{eleid}, 
and muons were identified using 
the two methods described in ref.~\cite{muid}.
The track momentum of the electron or muon candidate
was required to be larger than 2~GeV\@.
A jet reconstructed in cut (B-H3) was identified as a tau decay
if it contained only one or three charged tracks,
the invariant mass of the charged particles in the jet was
smaller than 1.5~GeV,
the invariant mass including energies deposited in the 
calorimeters was smaller than 2~GeV and
the scalar sum of momenta of the charged tracks was larger than 2~GeV.
 
\item[(B-H6)] 
The invariant mass excluding the most energetic lepton
was required to be smaller than 60~GeV
in order to reject $\WW \ra \nulqq $ events. 
As shown in Fig.~7, a large fraction of four-fermion
events was rejected using this requirement.
  
\item[(B-H7)] 
The visible mass of the events, $\Mvis$, must be smaller than 80~GeV 
to reduce $\WW$ background events in which 
one of $\Wpm$'s decayed into $\tau \nu$ and the other 
into ${\rm q \bar{q}^{'}(g)}$\@.
If one jet from ${\rm q \bar{q}^{'}(g)}$ was misidentified as tau lepton,
this event could pass through the previous cut (B-H6).
Such events were rejected by this requirement.   
Fig.~8 shows the distribution of  $\Mvis$.

\end{description}

No events were observed in the data after the above cuts.
The number of expected background events was 2.1\@.
Uncertainties of the expected background events
will be discussed in section~4.2\@.
The detection efficiencies for $\stoppair$ events
are listed in Table~\ref{tab:effbln}\@.
The efficiencies of both selection criteria (B-L and B-H)
are presented in this table.
As shown in this table, the detection efficiencies
for $\stopm \ra \bq \tau^{+} \snu_{\tau}$ are slightly smaller than
the case in which the branching fraction to each lepton flavour is assumed
to be the same.
 
\begin{table}[h]%------------------------------------------------------
\centering
\begin{tabular}{|l||r||r||r|r|r|r||r|r|r|}
\hline%-----------------------------------------------------------------
   &  \multicolumn{1}{c||}{data}         & \multicolumn{1}{c||}{total} 
   &  \multicolumn{1}{c|}{$\qq (\gamma)$} 
   &  \multicolumn{1}{c|}{$\ellell (\gamma)$} 
   &  \multicolumn{1}{c|}{`$\gamma \gamma$'} 
   &  \multicolumn{1}{c||}{4-f} 
   &  \multicolumn{3}{c|}{\rule{0mm}{6mm} $\stoppair$}  \\
   &       & \multicolumn{1}{c||}{bkg.} 
   &  &  &  &  & \multicolumn{3}{c|}{ }   \\
\hline %-----------------------------------------------------------------
$m_{\stopm}$ (GeV)&       &      &         &          &            &
        & \multicolumn{1}{c|}{80} & 
          \multicolumn{1}{c|}{80} &
          \multicolumn{1}{c|}{80}  \\
%------------------------------------------------------------------
$m_{\snu}$ (GeV)&       &      &         &            &            &
        & \multicolumn{1}{c|}{70} & 
          \multicolumn{1}{c|}{60} & 
          \multicolumn{1}{c|}{40}  \\
\hline %------------------------------------------------------------
 cut (B-H1)
& 22927 & 19603 &  1127 &  236 &  17670 &  567 & 856 & 792 & 784 \\
\hline% ----------------------------------------
 cut (B-H2)
&  1360 &  1381 &   724 &  151 &   32.2 &  474 & 490 & 748 & 776 \\
\hline% ----------------------------------------
 cut (B-H3)
&  857  &  915  &   466 & 1.48 &   8.66 &  438 & 490 & 743 & 764 \\
\hline% ----------------------------------------
 cut (B-H4)
&  209  &  213  &  15.3 & 0.36 &  2.60  &  195 & 424 & 665 & 660 \\
\hline% ----------------------------------------
 cut (B-H5)
&  162  &  162  &  6.59 & 0.33 & 0.32   &  155 & 369 & 639 & 637 \\
\hline% ----------------------------------------
 cut (B-H6)
&    4  &  6.35 &  0.17 & 0.14 & 0.22   & 5.83 & 369 & 639 & 606 \\
\hline% ----------------------------------------
 cut (B-H7)
&    0  & 2.05  &  0.15 &  0.06 & 0.22   & 1.62 & 369 & 639 & 581 \\ 
\hline% ----------------------------------------
\end{tabular}
\caption[]{
The remaining numbers of events
normalised to the
integrated luminosity of the data
for various background processes are compared with data
after each cut.
Numbers for
three simulated event samples of $\stoppair$ are 
also given (each starting from 1000 events).
In these samples, the branching fraction to each lepton flavour
is assumed to be the same.
Before the cut (B-H2),
there is a discrepancy between data and the simulated background processes,
since the numbers of events expected from two-photon processes
do not include the region
$M_{\gamma \gamma} <5$~GeV\@.
}
\label{tab:nevBH}
\end{table}%------------------------------------------------------------
 
\begin{table}[h]%------------------------------------------------------
\centering
\begin{tabular}{|l||r|r|r|r|r|r|r|r|}
\hline%-----------------------------------------------------------------
\multicolumn{9}{|c|}{\rule{0mm}{6mm} $\stopm \ra \bq \ell \snu$} \\ 
\multicolumn{9}{|c|}{the equal branching fractions 
for $\ell$= e, $\mu$, $\tau$} \\
\hline%-----------------------------------------------------------------
$m_{\stopm}$ (GeV) &  50 &  55 &  60 &  65 &  70 &  75 &  80 &  85 \\
\hline%-----------------------------------------------------------------
\hline%-----------------------------------------------------------------
$\Delta m$ (B-L)        &     &     &     &     &     &     &     &    \\
\hline%-----------------------------------------------------------------
 7~GeV                  &   9 &  11 &  12 &  12 &  12 &  12 &  12 &   9 \\
\hline%-----------------------------------------------------------------
10~GeV                  &  16 &  22 &  30 &  32 &  35 &  39 &  38 &  37 \\
\hline%-----------------------------------------------------------------
\hline%-----------------------------------------------------------------
$\Delta m$ (B-H)        &     &     &     &     &     &     &     &     \\
\hline%-----------------------------------------------------------------
10~GeV                  &  30 &  33 &  36 &  37 &  36 &  40 &  37 &  35 \\
\hline%-----------------------------------------------------------------
20~GeV                  &  45 &  52 &  56 &  60 &  61 &  63 &  64 &  65 \\
\hline%-----------------------------------------------------------------
$\mstop$/2              &  41 &  46 &  48 &  52 &  50 &  51 &  58 &  60 \\
\hline%-----------------------------------------------------------------
$\mstop$--10~GeV        &  25 &  26 &  27 &  27 &  29 &  26 &  28 &  27 \\
\hline%-----------------------------------------------------------------
\end{tabular}

\smallskip

\begin{tabular}{|l||r|r|r|r|r|r|r|r|}
\hline%-----------------------------------------------------------------
\multicolumn{9}{|c|}{\rule{0mm}{6mm} $\stopm \ra \bq \tau \snu_{\tau}$,
100\% branching fraction} \\
\hline%-----------------------------------------------------------------
$m_{\stopm}$ (GeV) &  50 &  55 &  60 &  65 &  70 &  75 &  80 &  85 \\
\hline%-----------------------------------------------------------------
\hline%-----------------------------------------------------------------
$\Delta m$ (B-L)        &     &     &     &     &     &     &     &     \\
\hline%-----------------------------------------------------------------
 7~GeV                  &   6 &   6 &   8 &   8 &   7 &   6 &   6 &   4 \\
\hline%-----------------------------------------------------------------
10~GeV                  &  16 &  22 &  26 &  29 &  31 &  32 &  29 &  28 \\
\hline%-----------------------------------------------------------------
\hline%-----------------------------------------------------------------
$\Delta m$ (B-H)        &     &     &     &     &     &     &     &     \\
\hline%-----------------------------------------------------------------
10~GeV                  &  19 &  20 &  21 &  21 &  22 &  23 &  22 &  19 \\
\hline%-----------------------------------------------------------------
20~GeV                  &  36 &  40 &  42 &  46 &  49 &  50 &  51 &  52 \\
\hline%-----------------------------------------------------------------
$\mstop$/2              &  35 &  38 &  44 &  44 &  48 &  50 &  54 &  51 \\
\hline%-----------------------------------------------------------------
$\mstop$--10~GeV        &  26 &  29 &  31 &  29 &  33 &  32 &  32 &  35 \\
\hline%-----------------------------------------------------------------
\end{tabular}
\caption[]{
The detection efficiencies in percent for $\stoppair$, 
in which $\stopm$ decays into $\bq \ell \snu$ 
($\ell = {\mathrm e},\mu,\tau$)\@.
The upper half of the table shows the case in which
the branching fraction to each lepton flavour is the same
and the lower half shows the worst case in which 
the branching fraction of $\stopm \ra  \bq \tau \snu_{\tau}$ 
is 100\%\@.
In both tables, $\Delta m$ is defined as $m_{\stopm}-m_{\snu}$\@.
The efficiencies in the first two lines of each table 
were obtained using the analysis B-L, the the
last four lines using analysis B-H\@. 
}
\label{tab:effbln}
\end{table}%------------------------------------------------------------

%=======================================================================
\section{Results}
%=======================================================================

No evidence for $\stoppair$ and $\sbotpair$ pair-production 
has been observed in the data. 
The data are consistent with the expected background 
of 1.9 events in analysis A, and 1.0 and 2.0 events\footnote{These
numbers were corrected for the inefficiency due to beam-related
background events. } in analysis B 
for an integrated luminosity of 56.8~$\pbinv$.
The sum of the expected number of background is 4.5 events 
subtracting overlap between three analyses.
Uncertainties of expected background will be discussed in section~4.2\@.
%Lower limits on $\mstop$ and $\msbot$ were calculated.
%The results obtained at lower centre-of-mass energies ($\rs$ = 130, 136, 161 
%and 171~GeV )~\cite{stop161,stop171} 
%were included in calculating these limits.
 
%=======================================================================
\subsection{Systematic Errors in the number of expected signal events}
%=======================================================================

The following sources of systematic error on the expected number 
of the signal events were taken into account:
 
\begin{enumerate}

\item The statistical error of the signal Monte Carlo simulation is
2--10\% depending on detection efficiencies.\\

\item The dependence of the detection efficiency on the mixing angle: \\ 
The energy distribution of the initial-state radiation  
depends on the mixing angle of the $\stopm$ ($\sbotm$),
because it influences 
the coupling between the $\stopm $ ($\sbotm$) and the $\Zboson$\@.
When the coupling is large, the initial-state radiation is hard.  
The detection efficiencies therefore depend on $\mixstop$ ($\mixsbot$)\@.
However, the detection efficiencies 
in Tables~\ref{tab:effcn}, \ref{tab:effbn} and \ref{tab:effbln} 
were calculated using the simulated events which were generated for 
$\mixstop$ = $\mixsbot$ = 0.0\@.

The detection efficiencies 
in the two extreme cases of $\stopm$ decoupled 
from the $\Zboson$ ($\mixstop = 0.98$) and  
$\stopm$ = $\stopl$ ($\mixstop$ = 0.0) were compared
for various $\mstop$ values.
The difference was always found to be 
within 2\%\@.
The effect on efficiencies 
for $\stopm \ra \bq \ell \snu$ and $\sbotm \ra \bq \neutralino$
was also checked and similar results were obtained. 
The systematic error due to the dependence on the mixing angle
was taken to be 2\%\@.
 
\item Fragmentation function for $ \stopm $: \\
The fragmentation scheme proposed by Peterson \etal $\,$ was used,
with the fragmentation parameter $ \epsilon_{\stopm} $ determined
by equation (1)\@.
The error on $ \epsilon_{\stopm} $ was propagated from
${\delta \epsilon_{\bq}}/{\epsilon_{\bq}} = \pm $ 
0.26~\cite{opalfragment} 
and ${\delta m_{\bq}}/{m_{\bq}} = \pm$ 0.06~\cite{PDG},
corresponding to $ \delta \epsilon_{\stopm} / \epsilon_{\stopm} 
 = \pm$ 0.27\@.
The systematic error in the efficiencies due to this uncertainty
was evaluated by altering the $\epsilon_{\stopm}$ parameter 
by one standard deviation for several combinations 
of ($m_{\stopm}$, $m_{\neutralino}$) and 
($m_{\stopm}$, $m_{\snu}$).
For the $\stopm \ra \cq \neutralino$ mode, 
the detection efficiencies changed by no more than
5\% over the $\mstop$ range.
The relative changes for the $\stopm \ra \bq \ell \snu$ mode
were found to be 6--10\%, and they depended mainly on $\mstop$.

To estimate the dependence on the fragmentation scheme,
the fragmentation function proposed by Bowler~\cite{Bowler} was used, 
because the shape of this fragmentation
function is very different from that of the Peterson function. 
The relative difference in efficiencies between 
the two fragmentation models
was 2--3\% for the $\stopm \ra \cq \neutralino$ mode, 
which was smaller than that due to
the variation of the $\epsilon_{\stopm}$ parameter used in 
the Peterson fragmentation scheme.
The systematic error due to the dependence on the fragmentation model
was taken to be 3\%\@.
For the $\stopm \ra \bq \ell \snu$ mode, the relative difference in 
efficiencies was found to lie between 4--8\%, where the range was
mainly due to $\mstop$\@.
 
\item Fragmentation function for $ \sbotm $: \\
The error due to the fragmentation function for $\sbotm$ was also 
estimated using the methods described above.
The uncertainty in $ \epsilon_{\sbotm} $ made a relative difference
of 4--6\% in the efficiencies. 

\item Fragmentation of the charm and bottom quarks:\\
The error in the efficiencies for the $\stopm \ra \cq \neutralino $ mode, 
due to the uncertainty in $ \epsilon_{\cq} $, was estimated to be typically 3\%
by changing $ \epsilon_{\cq} $ by 
$\delta \epsilon_{\cq} / \epsilon_{\cq} = \pm $0.35~\cite{opalfragment}.
 
The uncertainty in the $\epsilon_{\bq}$ parameter also contributes to 
the error in the efficiencies for the $\stopm \ra \bq \ell \snu$ and 
$\sbotm \ra \bq \neutralino$ modes.
As mentioned above, 
the $\epsilon_{\bq}$ parameter was simultaneously changed by $\pm$26\% 
when $\epsilon_{\stopm}$ and $\epsilon_{\sbotm}$ were altered.
The systematic error due to the uncertainty on $\epsilon_{\bq}$ 
is therefore taken into account in the errors  
$\delta \epsilon_{\stopm}$ and $\delta \epsilon_{\sbotm}$\@.

\item Fermi motion of the spectator quark in $\stopm$ 
($\sbotm$) -hadron decay:\\
Due to the Fermi motion of the spectator quark  
the invariant mass of the hadronic decay products of a 
$\stopm$ ($\sbotm$) -hadron varies.
For $\stopm \ra \cq \neutralino$ and $\sbotm \ra \bq \neutralino$ modes
this effect is not negligible when $\Delta m$ is large.
The systematic error in the efficiencies due to the Fermi motion
was evaluated by altering the mass of spectator quarks.
For the case of a 80~GeV $\stopm$ ($\sbotm$) and a massless neutralino
the efficiency varies by up to 6\% (8\%) due to the jet mass cut (A6).  
 
\item Lepton identification:\\
The systematic error on electron identification was estimated to be
4\% and the error on muon identification was 2\%\@.
The systematic error on tau identification is dominated by the 
uncertainties in the fragmentation of the bottom quark,
which has already been included in the uncertainty
in the $\epsilon_{\bq}$ parameter. 
A conservative error of 4\% was applied for all types of leptons.

\item Systematic errors due to imperfections in the 
Monte Carlo simulation of $P_t$, $P_t^{\mathrm{HCAL}}$,
the number of reconstructed jets, 
$\Evis$ and $\Mvis$ were estimated to be 3\%\@.

\item The integrated luminosity was calculated using the SW detector.
The systematic error on this luminosity was determined to be 
0.26\% (stat.) and 0.41\% (syst.)\@.
 
\item The systematic error due to the uncertainty on the trigger
efficiency was estimated to be negligible.
This is expected because of the requirement of at least four good tracks.
 
\end {enumerate}

The various systematic errors are summarised in Table~\ref{sys3}\@.
These systematic errors were considered to be independent and
the total systematic error was calculated as the quadratic sum
of the individual errors.
These systematic errors were treated as in Ref.~\cite{SYSTEM} in 
calculating the limits.

\begin{table}[h]%------------------------------------------------------
\centering
\begin{tabular}{|r|r|r|r|} \hline
Sources & \multicolumn{1}{|c|}{\rule{0mm}{6mm} $\stopm \ra \cq \neutralino$} 
        & \multicolumn{1}{|c|}{\rule{0mm}{6mm} $\stopm \ra \bq \ell \snu $}
        & \multicolumn{1}{|c|}{\rule{0mm}{6mm} $\sbotm \ra \bq \neutralino $}
\\ \hline \hline
Statistical error of MC & \multicolumn{3}{|c|}{2--10\%} \\ \hline 
% $ \frac{\Delta \eta }{\eta } = \sqrt{(1-\frac{n}{N_0}) / n} $} \\ \hline
$\mixstop$ dependence & 2\% &  2\% &  --   \\ \hline
$\mixsbot$ dependence & --    &  --    & 2--4\% \\ \hline
Uncertainty on $\epsilon_{\stopm}$  & 5\%   & 6--10\% &   -- \\ \hline
Uncertainty on $\epsilon_{\sbotm}$  & --    & --    & 4--6\% \\ \hline
Fragmentation scheme & 3\%  &  4--8\% & 4--10\% \\ \hline
Uncertainty on $\epsilon_{\cq}$ & 3\%  &  -- & -- \\ \hline
Uncertainty on $\epsilon_{\bq}$ & --  & 
         \multicolumn{2}{|c|}{Included in the } \\
 & & \multicolumn{2}{|c|}{ uncertainties of $\epsilon_{\stopm}$ 
              and $\epsilon_{\sbotm}$}  \\ \hline
Spectator Fermi motion & 3--6\%  &  4\%  & 3--8\% \\ \hline
Uncertainty of lepton ID & -- & 4\% & -- \\ \hline
Detector simulation &  \multicolumn{3}{|c|}{ 3\% }  \\ \hline
Luminosity  &  \multicolumn{3}{|c|}{ 0.5\% }  \\ \hline
Trigger efficiencies &  \multicolumn{3}{|c|}{ negligibly small }  \\ \hline
\end{tabular}
\caption[]{
The summary of the systematic errors on the expected number of the 
signal events. 
The range of these errors depend on 
the mass of $\stopm$ and $\sbotm$\@.
}
\label{sys3}
\end{table}%------------------------------------------------------------

%=======================================================================
\subsection{Systematic Errors in the number of background events}
%=======================================================================

The two-photon and four-fermion processes are the dominant background.
Systematic errors in the expected number of 
these processes are discussed here.
Since no event was observed in data,
the expected number of background events was not 
subtracted to calculate limits. 
Then these errors were not used to calculate limits.

\subsubsection{Two-photon processes}

The systematic errors are mostly dominated 
by the Monte Carlo statistics for two-photon processes.
The statistical fluctuations of the expected numbers
(relative errors)  
are $\pm$0.23(85\%), $\pm$0.49(50\%) 
and $\pm$0.22(100\%) for analysis A, B-L and B-H, respectively.

Furthermore the uncertainty on the modelling of 
the two-photon processes was checked with data.
In order to select two-photon events the visible energy was required to be 
smaller than 20\% of $\sqrt{s}$, the charged multiplicity to be 
at least four, the visible invariant mass to be larger than 5 GeV
and the forward detector vetoes (cut P2) were required. 
The $P_t$ distributions of the selected events from data were
compared with Monte Carlo.
The shapes of the distributions agree with each other,
but there is an uncertainty of 30\% in the normalisation.

\subsubsection{Four-fermion processes}

Uncertainties in the generators of the four-fermion processes 
were estimated by comparing grc4f with the 
Excalibur~\cite{excalibur} and PYTHIA~\cite{lund} generators.
The background events predicted by these different generators are
summarised in Table~8\@.
The differences larger than statistical fluctuations 
were found especially in analysis A.
The difference between grc4f and Excalibur comes mainly from 
the region of $M_{\qqx } < $ 40~GeV 
for $\ee \ra {\rm e} \nu \qqx $ process.
On the other hand, the difference between grc4f and PYTHIA
comes from the region of $M_{\qq} >$ 15~GeV 
for $\ee \ra \gamma ^{*} \Zboson \ra \qq \nunu $ process.
The prediction on this process by PYTHIA is 
about 60\% of grc4f.   
These differences were considered as the systematic errors 
in the four-fermion processes;
$1.63 \pm 0.13$ (stat.) $^{+0.52}_{-0.33}$ (sys.) for analysis A,
$0.09 \pm 0.03$ (stat.) $^{+0.12}_{-0.06}$ (sys.) for analysis B-L
and
$1.62 \pm 0.13$ (stat.) $^{+0.23}_{-0.18}$ (sys.) for analysis B-H\@.

\begin{table}[h]%------------------------------------------------------
\centering
\begin{tabular}{|l||r||r|r|}
\hline%-----------------------------------------------------------------
&  \multicolumn{1}{c||}{grc4f}
&  \multicolumn{1}{c|}{Excalibur}
&  \multicolumn{1}{c|}{PYTHIA} \\
\hline%-----------------------------------------------------------------
\hline%-----------------------------------------------------------------
analysis A   & $1.63 \pm 0.13$ & $2.15 \pm 0.15$ & $1.30 \pm 0.10$ \\
\hline%-----------------------------------------------------------------
analysis B-L & $0.09 \pm 0.03$ & $0.21 \pm 0.05$ & $0.03 \pm 0.01$ \\
\hline%-----------------------------------------------------------------
analysis B-H & $1.62 \pm 0.13$ & $1.85 \pm 0.13$ & $1.44 \pm 0.13$ \\
\hline%-----------------------------------------------------------------
\end{tabular}
\caption[]{
The expected numbers of four-fermion background processes 
predicted by different three generators.
The errors shown in this table are the statistical fluctuations.
}
\label{tab:bg}
\end{table}%------------------------------------------------------------

Events with two jets plus missing transverse momentum
were checked with data to study the differences in analysis A\@.
To select events with two jets and large $P_t$ coming from
the four-fermion processes,  
the cuts P1, P2 and A1 were applied, 
and the $P_t$ was required to be greater than 10~GeV
to reject two-photon processes completely.
Furthermore $d_{23}^{2} (\equiv y_{23} {\Evis}^{2})$ was required to be
smaller than 50~GeV$^{2}$ to select clear two-jet events,
where $y_{23}$ is the jet resolution parameter from 2 jets to 3 jets 
using the Durham algorithm.
Finally, the acoplanarity angle, $\phiacop$, was required to be larger 
than 10$\degree$ or 100$\degree$\@.
After these selections, the observed numbers in data were compared
with the predictions by these three generators (Table~9)\@.

\begin{table}[h]%------------------------------------------------------
\centering
\begin{tabular}{|l|r||r|r|r|}
\hline%-----------------------------------------------------------------
&  \multicolumn{1}{c||}{DATA}
&  \multicolumn{1}{c|}{grc4f}
&  \multicolumn{1}{c|}{Excalibur}
&  \multicolumn{1}{c|}{PYTHIA} \\
\hline%-----------------------------------------------------------------
\hline%-----------------------------------------------------------------
$\phiacop > 10\degree$ & 19
             & $17.2 \pm 0.4$ & $18.9 \pm 0.4$ &  $17.7 \pm 0.1$ \\
\hline%-----------------------------------------------------------------
$\phiacop > 100 \degree$ & 2
             & $1.87 \pm 0.15$ & $3.25 \pm 0.18$ &  $1.22 \pm 0.09$ \\
\hline%-----------------------------------------------------------------
\end{tabular}
\caption[]{
The remaining numbers of events in data and three different 
generators.
The errors shown in this table are the statistical errors.
}
\label{tab:bg2}
\end{table}%------------------------------------------------------------

In the region of $\phiacop > 100 \degree$,  
large differences were observed in the predictions 
of these three generators.
But all three predictions are consistent with the data,
since the statistics of data is too small.
We need data with higher statistics to study the four-fermion generators.

%=======================================================================
\subsection{Mass Limits}
%=======================================================================

\subsubsection{Scalar top quark {\boldmath$\boldsymbol{\stopm}$}} 

To calculate mass limits, the number of signal events
passing through the event selections is determined as a function of 
$\mstop$, $\mchi$ (or $m_{\snu}$) and $\mixstop$\@.

Figs.~9(a), 10(a) and 11(a) show the 95\% C.L. excluded regions
in the ($\mixstop$ , $m_{\stopm}$) plane for 
the $\stopm \ra \cq \neutralino$, $\stopm \ra \bq \ell \snu$
($\ell$= e,$\mu$,$\tau$) and $\stopm \ra \bq \tau \snu$ decay modes, 
respectively.
The branching fraction to each lepton flavour $\ell^{+}$ depends
on the composition of the lightest chargino~\cite{stop171}.
As the chargino becomes Higgsino-like, the branching fraction into 
$\bq \tau ^{+} \snu_{\tau} $ becomes large.
In the limit that the chargino is the pure Wino state,
the branching fraction to each lepton flavour is the same.
Two extreme cases in which
the branching fraction to each lepton flavour is the same, or
the branching fraction into $\bq \tau ^{+} \snu_{\tau} $ is 100\%, 
were considered here\@.
The 95\% C.L. mass bounds are listed in Table~\ref{limit1}
for various values of $\mixstop$\@.
Assuming that the $\stopm$ decays into $\cq \neutralino$,
and that the mass difference between the $\stopm$ and the $\neutralino$
is greater than 10~GeV,
the $\stopm$ is found to be heavier than 85.0~GeV,
if $\mixstop$ = 0\@. 
A lower limit of 81.3~GeV is obtained
even if the $\stopm$ decouples from the $\Zboson$ boson. 
When the $\stopm$ decays into $\bq \ell \snu$,
the lower limit on $\mstop$ is 83.6~GeV, 
assuming that the mass difference between $\stopm$ and $\snu$
is greater than 10~GeV, that $\mixstop$ = 0 
and that the branching fraction to each lepton flavour is the same.
The 95\% C.L. excluded regions in the ($\mstop$ , $\mchi$) and 
($\mstop$ , $m_{\snu}$) planes 
are shown in Figs.~9(b), 10(b) and 11(b) 
for various values of $\mixstop$\@.
In these figures, the limits calculated by the expected 
background events are superimposed to show the sensitivity of
these analyses.

\begin{table}[h] \centering
\begin{tabular}{|r || c | c || c | c | } \hline
\multicolumn{5}{|c|}{ \rule{0mm}{6mm} 
            Lower limit on $\mstop$ (GeV) } \\ \hline 
 & \multicolumn{2}{|c||}{ \rule{0mm}{6mm} $\stopm \ra \cq \neutralino$ }
 & \multicolumn{1}{|c|}{$\stopm \ra \bq \ell \snu$} 
 & \multicolumn{1}{|c|}{$\stopm \ra \bq \tau \snu_{\tau}$} \\ 
 & \multicolumn{2}{|c||}{ } 
 & \multicolumn{1}{|c|}{$\ell = {\mathrm e}, \mu, \tau$} 
 & \multicolumn{1}{|c|}{Br = 100\%} \\ \hline
 
$\mixstop$ (rad) & $ \Delta m \geq 5$~GeV & 
                 $ \Delta m \geq 10$~GeV  & $ \Delta m \geq 10$~GeV 
                 & $ \Delta m \geq 10$~GeV \\ \hline
0.0   &                           81.2 &  85.0 & 83.6 & 80.0  \\ \hline
$ \leq \frac{1}{8} \pi $ &        80.0 &  84.2 & 82.5 & 79.0  \\ \hline
$ \leq \frac{1}{4} \pi $ &        76.8 &  82.0 & 79.7 & 76.1  \\ \hline
 0.98 &                           75.8 &  81.3 & 79.2 & 75.0  \\ \hline
\end{tabular}
\caption[]{
The excluded $\mstop$ region at 95\% C.L. ($\Delta m  =  \mstop - \mchi$
 or $ \mstop - m_{\snu}$)\@. 
}
\label{limit1}
\end{table}
 
\subsubsection{Scalar bottom quark {\boldmath$\boldsymbol{\sbotm}$}} 

To calculate mass limits, the number of signal events
passing through the event selections is determined as a function of 
$m_{\sbotm}$, $\mchi$ and $\mixsbot$\@.
Fig.~12(a) shows the 95\% C.L. excluded regions
in the ($\mixsbot$, $m_{\sbotm}$) plane for the mass difference of
$\Delta m ( \equiv m_{\sbotm} - m_{\neutralino} ) \geq $ 7~GeV 
and $\Delta m \geq $ 10~GeV\@.
The numerical mass bounds are listed in Table~\ref{limit2}
for various $\mixsbot$\@.
The lower limit on the $\sbotm$-mass is 82.7~GeV,
if $\Delta m$ is greater than 7~GeV
and $\mixsbot$ = 0\@.
The 95\% C.L. excluded regions in the ($m_{\sbotm}$, $m_{\neutralino}$) 
plane are shown in Fig.~12(b) for various $\mixsbot$\@.
In this figure, the limit calculated by the expected 
background events is also superimposed to show the sensitivity of
this analysis.

\begin{table}[h] \centering
\begin{tabular}{|r || c | c |} \hline
\multicolumn{3}{|c|}{ \rule{0mm}{6mm} Lower limit on $m_{\sbotm}$ (GeV)
($\sbotm \ra \bq \neutralino$) } \\ \hline
 
$\mixsbot$ (rad) & $ \Delta m \geq 7$~GeV &  $\Delta m \geq 10$~GeV \\ 
                 &                  & $ m_{\neutralino} \geq 30$~GeV \\  \hline
0.0              &               82.7      &      84.0              \\ \hline
$ \leq \frac{1}{8} \pi $ &       81.0      &      82.6              \\ \hline
$ \leq \frac{1}{4} \pi $ &       71.9      &      76.2              \\ \hline
1.17                     &       54.4      &      63.7              \\ \hline
\end{tabular}
\caption[]{
The excluded $\msbot$ region at 95\% C.L. 
($\Delta m  = m_{\sbotm} - m_{\neutralino}$)
} 
\label{limit2}
\end{table}
 
%=======================================================================
\section{Summary and Conclusion}
%=======================================================================
 
A data sample collected using the OPAL detector
corresponding to an integrated luminosity of 56.8~pb$^{-1}$
at $\roots = $183~GeV has been analysed
to search for pair production of the scalar top quark and the 
scalar bottom quark predicted by supersymmetric theories.
R-parity was assumed to be conserved\@.
No events remained after the selection cuts.

The 95\% C.L. lower limits on the scalar top quark mass are 
85.0 and 82.0~GeV,  
if the mixing angle of the scalar top quark is 0 and 
smaller than $\frac{\pi}{4}$, respectively.
Even if the $\stopm$ decouples from the $\Zboson$ boson, 
a lower limit of 81.3~GeV is obtained.
These limits were obtained assuming that
the scalar top quark decays into a charm quark and the lightest neutralino 
and that the mass difference between the scalar top and the lightest 
neutralino is larger than 10~GeV\@.

Assuming a relatively light scalar neutrino
(37.1~GeV~\cite{snulimit,PDG} $ \leq m_{\snu} \leq \, \mstop - m_{\bq}$) 
the complementary decay mode of the scalar top quark 
in which it decays into a bottom quark, a charged lepton 
and the scalar neutrino has also been studied. 
If the mass difference between the scalar top quark and the scalar neutrino
is greater than 10~GeV
and if the mixing angle of the scalar top quark is 0,
the 95\% C.L. lower limit on the scalar top quark mass is 83.6~GeV\@.
This limit is obtained assuming that the branching fraction to
each lepton flavour is the same. 

A mass limit on the light scalar bottom quark is found to be
82.7~GeV (95\% C.L.),
assuming that the mass difference between the scalar bottom quark and 
the lightest neutralino is greater than 7~GeV
and that the mixing angle of the scalar bottom quark is zero.
If the mass difference is greater than 10~GeV
and the lightest neutralino is heavier than 30~GeV,
a mass limit on the light scalar bottom quark is 84.0~GeV (95\% C.L.).

%-----------------------------------------------------------------------
\section*{Acknowledgements}
%-----------------------------------------------------------------------

We particularly wish to thank the SL Division for the efficient operation
of the LEP accelerator at all energies
and for their continuing close cooperation with
our experimental group.  We thank our colleagues from CEA, DAPNIA/SPP,
CE-Saclay for their efforts over the years on the time-of-flight and trigger
systems which we continue to use.  In addition to the support staff at our own
institutions we are pleased to acknowledge the  \\
Department of Energy, USA, \\
National Science Foundation, USA, \\
Particle Physics and Astronomy Research Council, UK, \\
Natural Sciences and Engineering Research Council, Canada, \\
Israel Science Foundation, administered by the Israel
Academy of Science and Humanities, \\
Minerva Gesellschaft, \\
Benoziyo Center for High Energy Physics,\\
Japanese Ministry of Education, Science and Culture (the
Monbusho) and a grant under the Monbusho International
Science Research Program,\\
German Israeli Bi-national Science Foundation (GIF), \\
Bundesministerium f\"ur Bildung, Wissenschaft,
Forschung und Technologie, Germany, \\
National Research Council of Canada, \\
Research Corporation, USA,\\
Hungarian Foundation for Scientific Research, OTKA T-016660, 
T023793 and OTKA F-023259.\\

%=======================================================================
%       References
%=======================================================================

%=======================================================================
%       Figures
%=======================================================================
%-----------------------------------------------------------------------
\newpage 
%-----------------------------------------------------------------------
\begin{figure}[t]
\vspace*{-10.mm}
\begin{center}\mbox{
\epsfig{file=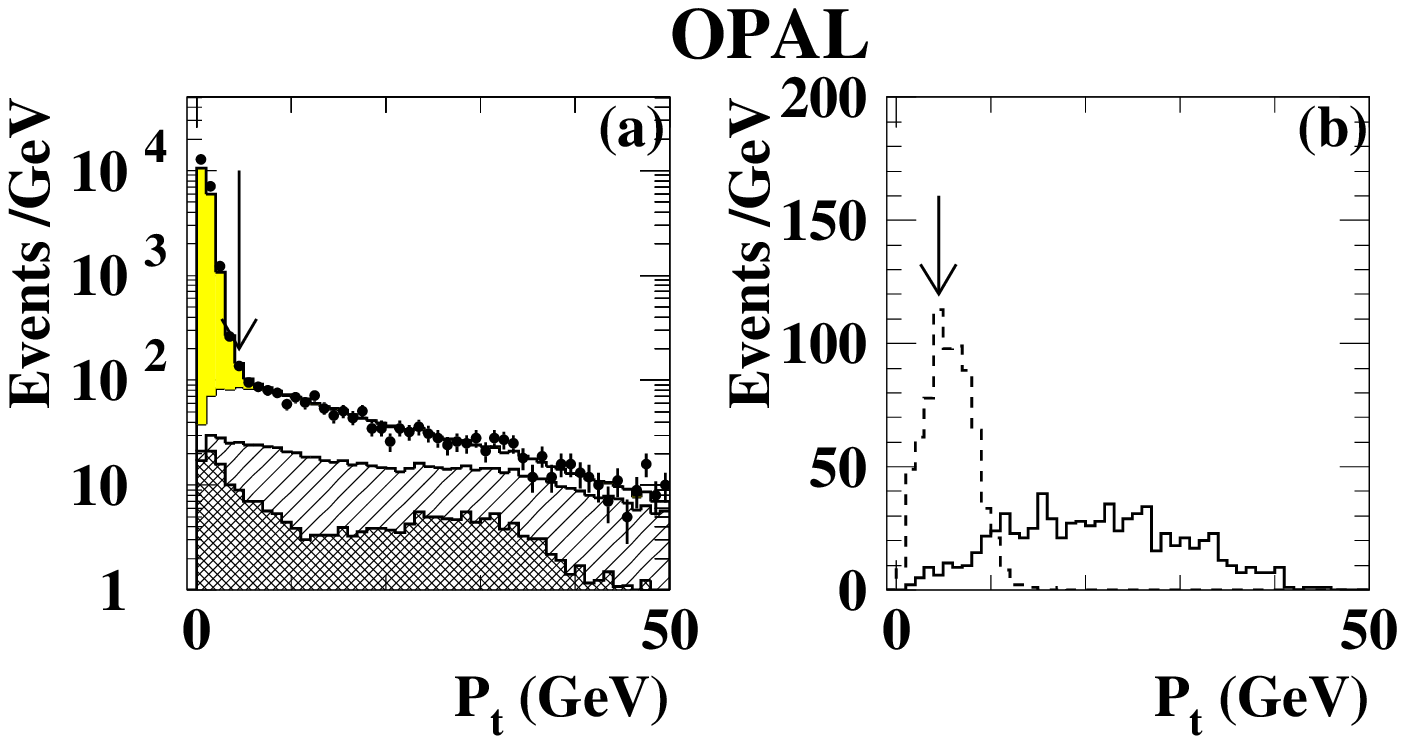,width=16.0cm}
}\end{center}
\vspace*{-10.mm}
\caption[]
{
The distributions of ${P}_{t}$ 
after cut (A1) 
for background (histograms) and data in (a), and for
$\stoppair$ predictions in (b)\@. The arrows in these figures show
the selection criteria. 
In (a) the distribution of the data is shown by the points 
with error bars.
The predictions from background processes are also shown: 
dilepton events (cross-hatched area), 
two-photon processes (grey area), 
four-fermion processes (singly-hatched area),
and multihadronic events (open area)\@.
(b) shows predictions for $\stoppair$ in which $\stopm$ decays
into $\cq \neutralino$.
The continuous line histogram is for ($\mstop$, $m_{\neutralino}$)
=(80~GeV, 60~GeV), and the dashed line is for (80~GeV, 75~GeV),
starting from 1000 generated events each.
}
\label{fig:ptfig}
%\end{figure}
%-----------------------------------------------------------------------
%\newpage
\nopagebreak [4] 
\vspace*{-10.mm}
%\begin{figure}[t]
\begin{center}\mbox{
\epsfig{file=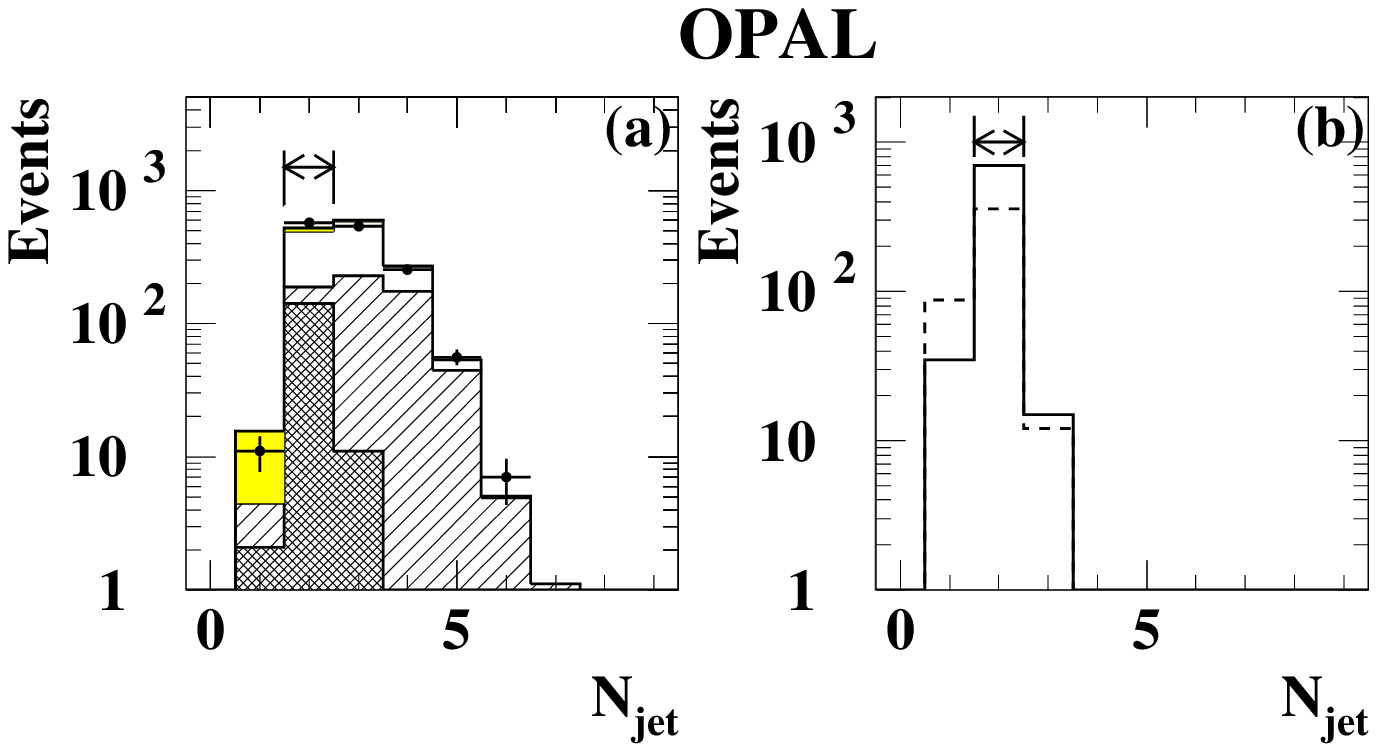,width=16.0cm}
}\end{center}
\vspace*{-10.mm}
\caption[]
{
The distributions of the number of reconstructed jets
after cut (A2):
for background (histograms) and data in (a), and for
$\stoppair$ predictions in (b).
The conventions for the various histograms are 
the same as in Fig. 1\@.
}
\label{fig:njetfig1}
\end{figure}
%-----------------------------------------------------------------------
\newpage
\begin{figure}[t]
\begin{center}\mbox{
\epsfig{file=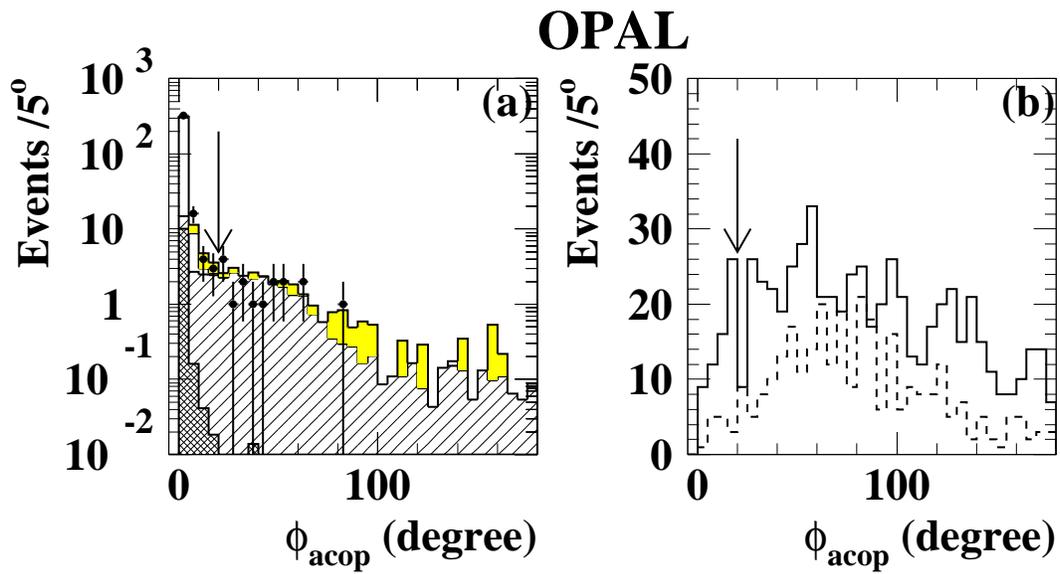,width=16.0cm}
}\end{center}
\caption[]
{
The distributions of the acoplanarity angle after cut (A3) for
background (histograms) and data in (a), 
and for the $\stoppair$ predictions in (b).
The conventions for the  
various histograms are the same as in Fig.~1.
}
\label{fig:acopfig1}
\end{figure}
%-----------------------------------------------------------------------
\newpage
\begin{figure}[t]
\begin{center}\mbox{
\epsfig{file=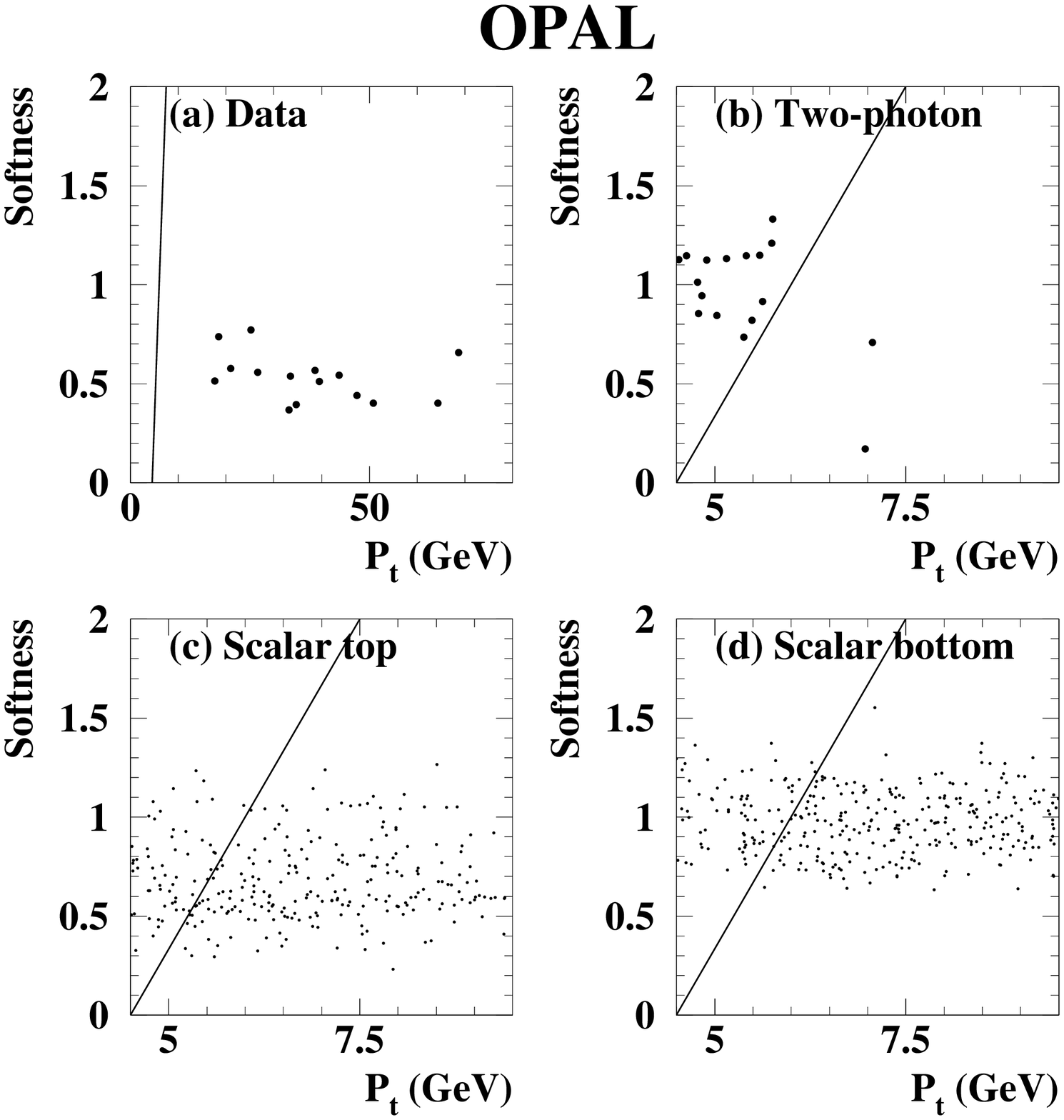,width=16.0cm}
}\end{center}
\caption[]
{
Scatter plots of `Softness'(see text) versus $P_t$
after cut (A4) for (a) data, 
(b) simulated two-photon processes,
(c) the Monte Carlo simulation of $\stoppair$ signals for  
($\mstop$,$m_{\neutralino}$)=(80~GeV, 75~GeV),
and (d) the Monte Carlo simulation of $\sbotpair$ signals for  
($\msbot$,$m_{\neutralino}$)=(80~GeV, 73~GeV).
The scale of $P_t$ in (a) is different from the other figures. 
The simulated events are not normalised to the luminosity.
For the two-photon processes, the corresponding luminosity is 
253~$\pbinv$. 
The event samples of $\stoppair$ and $\sbotpair$ each 
start from 1000 events.
}
\label{fig:mjetfig}
\end{figure}
%-----------------------------------------------------------------------
%-----------------------------------------------------------------------
\newpage
\begin{figure}[t]
\begin{center}\mbox{
\epsfig{file=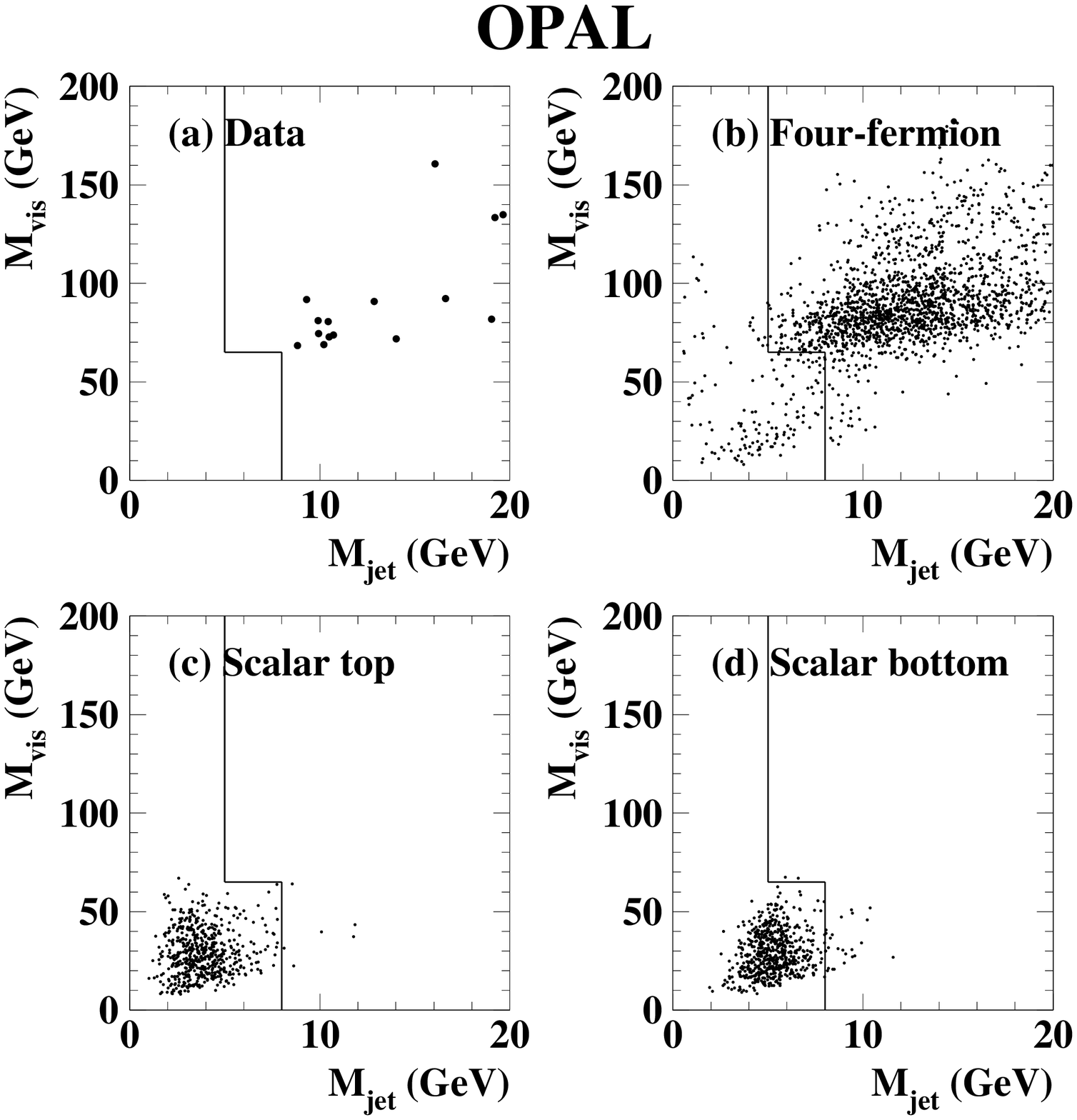,width=16.0cm}
}\end{center}
\caption[]
{
Scatter plots of $\Mvis$ versus $\mjet$ after cut (A5) for (a) data, 
(b) simulated four-fermion processes,
(c) the Monte Carlo simulation of $\stoppair$ signals for  
($\mstop$,$m_{\neutralino}$)=(80~GeV, 60~GeV),
and (d) the Monte Carlo simulation of $\sbotpair$ signals for  
($\msbot$,$m_{\neutralino}$)=(80~GeV, 60~GeV)\@.
The simulated events are not normalised to the luminosity.
For the four-fermion processes, the corresponding luminosity is 
5000~$\pbinv$. 
The event samples of $\stoppair$ and $\sbotpair$ each start from 1000 events.
}
\label{fig:ljetfig}
\end{figure}
%-----------------------------------------------------------------------
\newpage
\begin{figure}[t]
\vspace*{-10.mm}
\begin{center}\mbox{
\epsfig{file=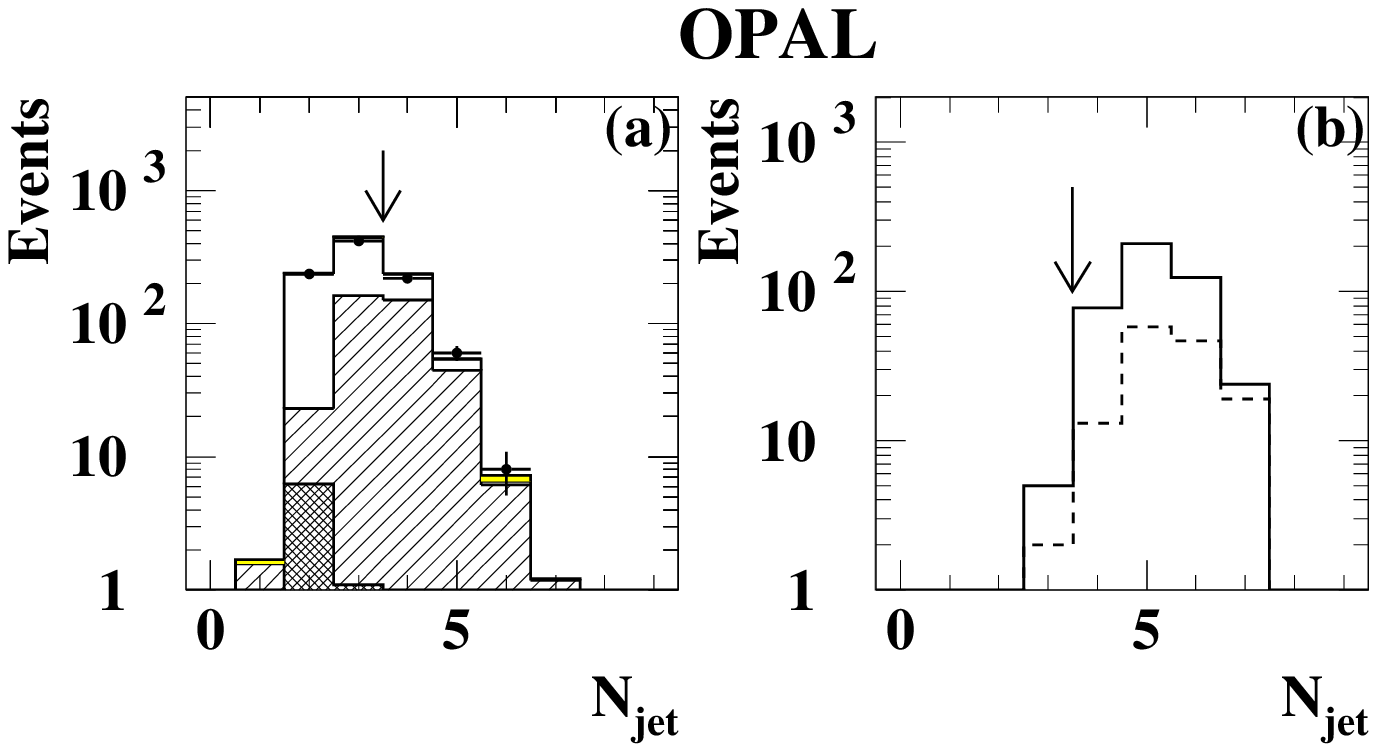,width=16.0cm}
}\end{center}
\vspace*{-10.mm}
\caption[]
{
The distributions of the number of reconstructed jets 
after cut (B-L2)\@.  The arrows in these figures show
the selection criteria.
(a) shows the distribution of the data with error bars.
The predictions from background processes are also shown: 
dilepton events (cross-hatched area), 
two-photon processes (grey area), 
four-fermion processes (singly-hatched area),
and the multihadronic events (open area).
(b) shows predictions for $\stoppair$ in which $\stopm$ decays
into $\bq \ell \snu$.
The continuous line histogram is for ($\mstop$, $m_{\snu}$)
=(80~GeV, 70~GeV), and the dotted line is for (80~GeV, 73~GeV)\@.
In these samples, the branching fraction to each lepton flavour
is assumed to be the same.
}
\label{fig:njetfig2}
%\end{figure}
%-----------------------------------------------------------------------
%\newpage 
%\begin{figure}[htb]
\vspace*{-15.mm}
\begin{center}\mbox{
\epsfig{file=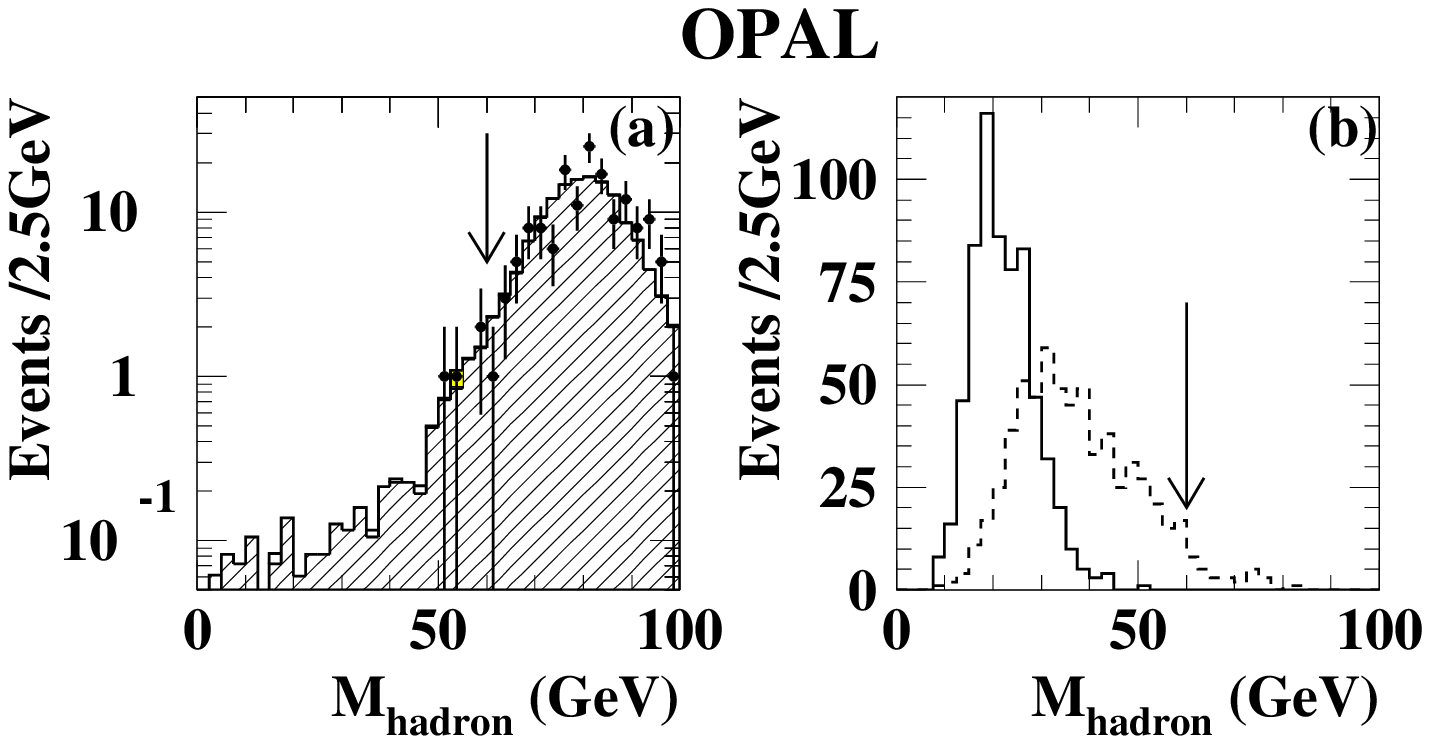,width=16.0cm}
}\end{center}
\vspace*{-13.mm}
\caption[]
{
The distributions of invariant mass excluding 
the most energetic lepton after cut (B-H5)\@.  
The arrows in these figures show
the selection criteria.
The conventions for the various histograms in (a) are 
the same as in Fig.~6.
In (b) the continuous line histogram is for ($\mstop$, $m_{\snu}$)
=(80~GeV, 60~GeV), and the dotted line is for (80~GeV, 40~GeV)\@.
In these samples, the branching fraction to each lepton flavour
is assumed to be the same.
}
\label{fig:mhad}
\end{figure}
%-----------------------------------------------------------------------
\newpage 
\begin{figure}[t]
\vspace*{-20.mm}
\begin{center}\mbox{
\epsfig{file=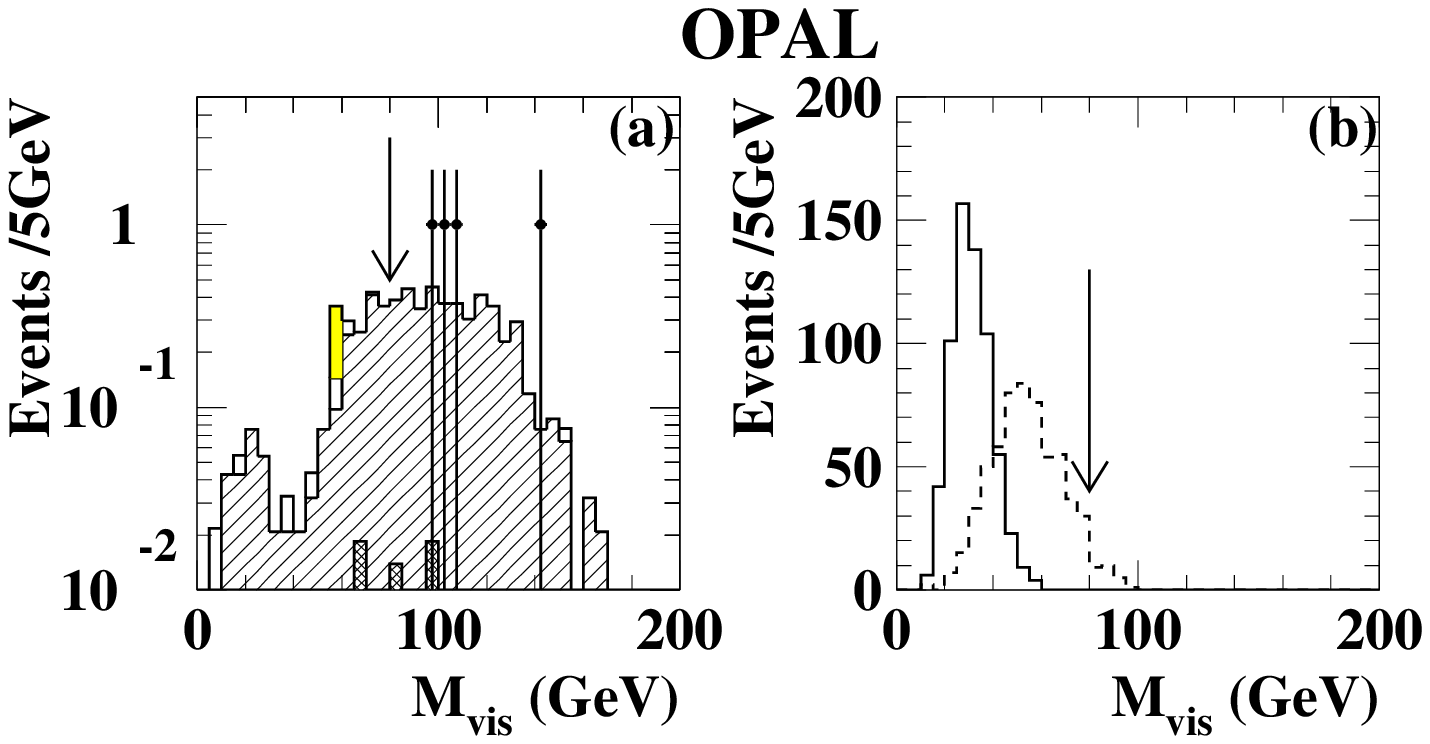,width=16.0cm}
}\end{center}
\caption[]
{
The distributions of the visible mass
after cut (B-H6) for
background (histograms) and data in (a), and for
the $\stoppair$ signal predictions in (b).
The conventions for the various histograms are the same as in Fig.~7.
}
\label{fig:evisfig2}
%\end{figure}
%-----------------------------------------------------------------------
%\newpage 
%\begin{figure}[htb]
\vspace*{-15.mm}
\begin{center}\mbox{
\epsfig{file=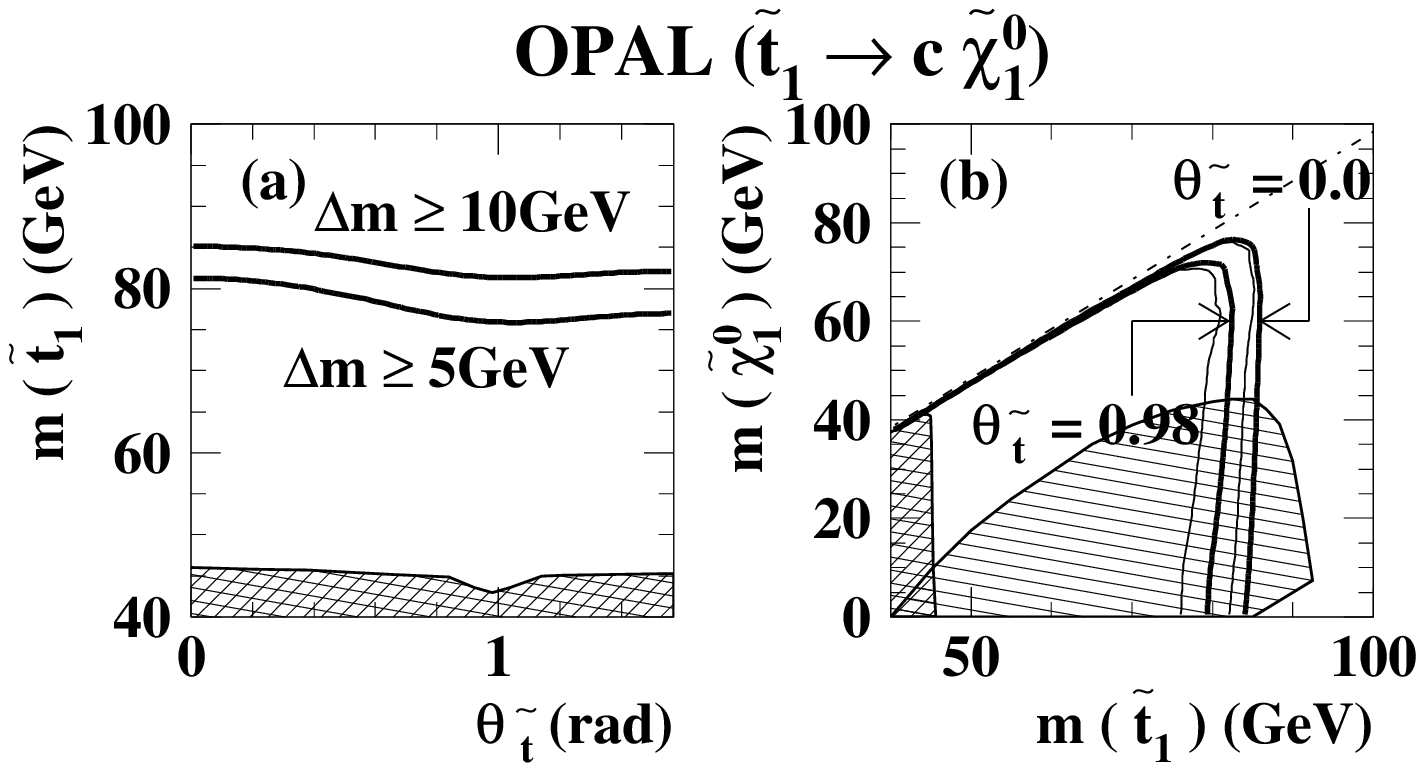,width=16.0cm}
}\end{center} 
\vspace*{-10.0mm}
\caption[]
{
The 95\% C.L. excluded regions 
assuming that the $\stopm$ decays into $\cq \neutralino$\@.
(a) The excluded regions in the $(\mixstop , \mstop)$ plane
for a mass difference 
$ \Delta m \, ( = \mstop - \mchi) \geq $ 10~GeV, 
and $ \Delta m \geq $ 5~GeV.
The cross-hatched region has already been excluded by the search at 
LEP1~\cite{opalstop}\@.
(b) The excluded regions in the $(\mstop, \, \mchi)$ plane,
for a mixing angle of $\stopm$ of 0.0 and 0.98 rad.
The solid lines show the actual limits, and 
the thin lines show the limits calculated only with the
expected number of background events. 
The cross-hatched region has already been excluded by the search 
at LEP1~\cite{opalstop}.
The singly-hatched region has been excluded 
by the D0 Collaboration~\cite{d0}.
The dash-dotted straight line shows the kinematic limit 
for the $\stopm \ra \cq \neutralino$ decay. 
}
\label{fig:result1}
\end{figure}
%-----------------------------------------------------------------------
\newpage 
\begin{figure}[htb]
\vspace*{-30.0mm}
\begin{center}\mbox{
\epsfig{file=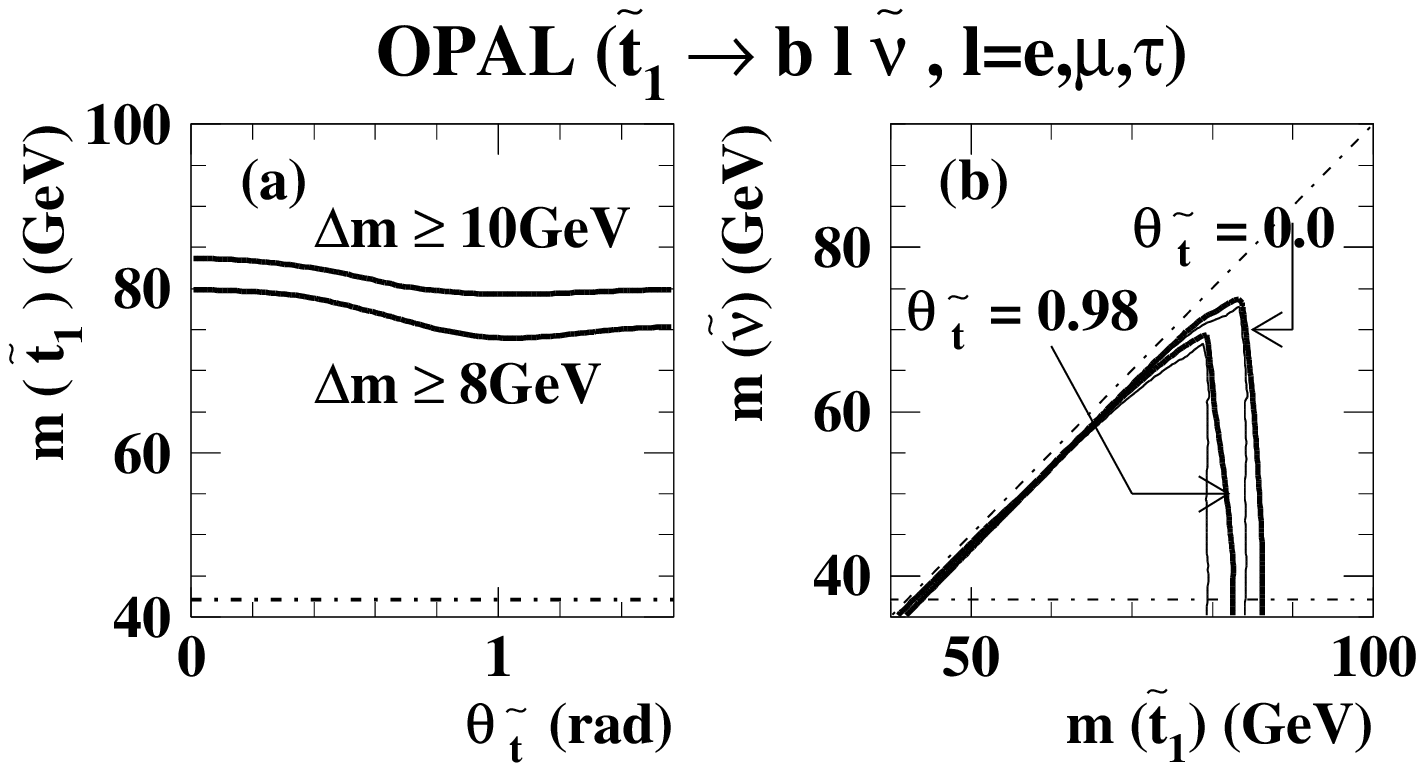,width=16.0cm}
}\end{center} 
\vspace*{-12.0mm}
\caption[]
{
The 95\% C.L. excluded regions 
assuming that the $\stopm$ decays into $\bq \ell \snu$
and that the branching fraction to each lepton flavour
is the same.  
(a) The excluded regions in the $(\mixstop, \, \mstop)$ plane
where the mass difference between the $\stopm$ and the $\snu$ 
is greater than 8 or 10~GeV\@.
The dash-dotted straight line shows the kinematic limit 
for this decay,
since a $\snu$ lighter than 37.1~GeV  
has been excluded~\cite{snulimit,PDG}\@.
(b) The excluded regions in the $(\mstop, \, m_{\snu})$ plane,
for a mixing angle of the $\stopm$ assumed to be
0.0 and 0.98 rad.
The solid lines show the actual limits, and 
the thin lines show the limits calculated only with the
expected number of background events. 
The dash-dotted horizontal line shows the limit on $m_{\snu}$
obtained at LEP1,
and the dash-dotted diagonal line shows the kinematic limit 
for the $\stopm \ra \bq \ell \snu$ decay. 
}
\label{fig:result2}
%\end{figure}
%-----------------------------------------------------------------------
%\newpage 
%\begin{figure}[htb]
\vspace*{-15.0mm}
\begin{center}\mbox{
\epsfig{file=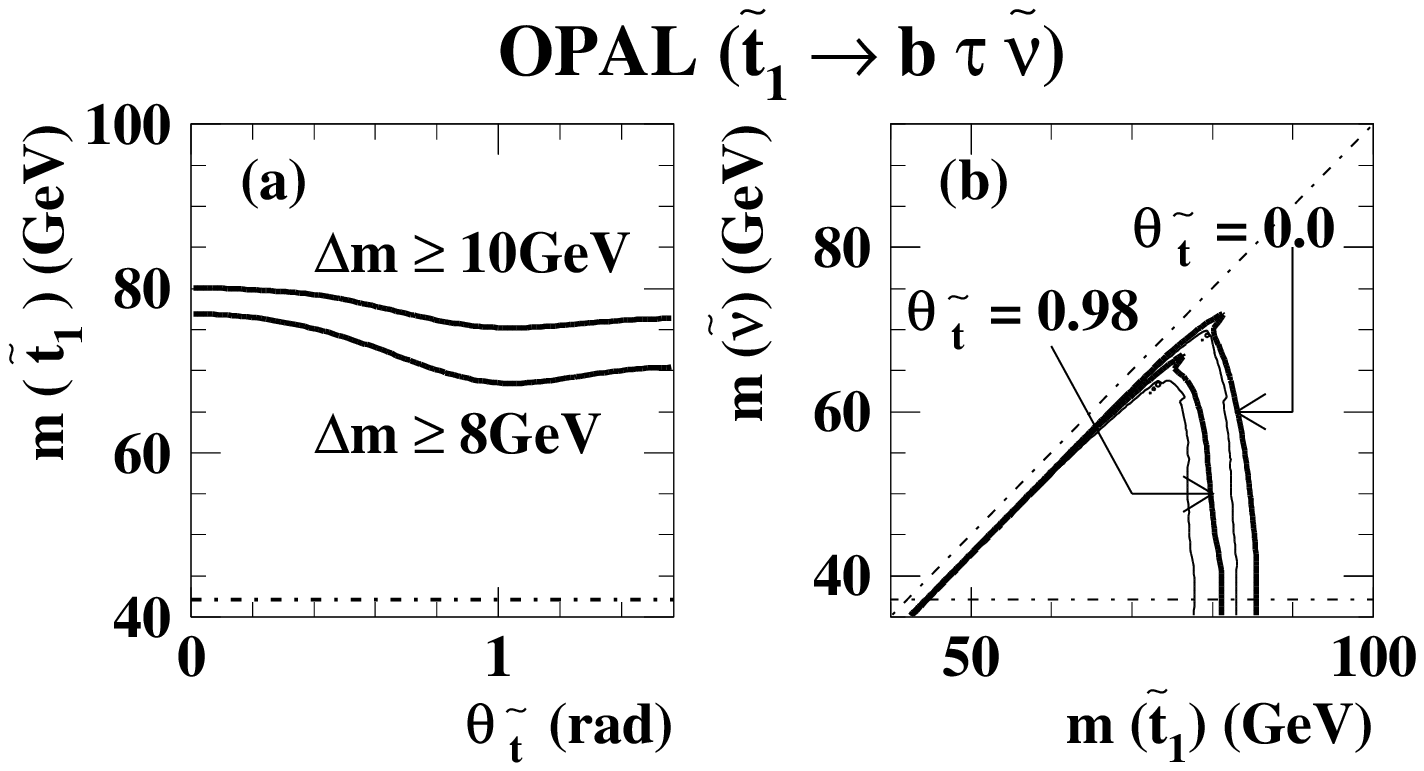,width=16.0cm}
}\end{center} 
\vspace*{-12.0mm}
\caption[]
{
The 95\% C.L. excluded regions 
assuming that $\stopm$ always decays into $\bq \tau \snu_{\tau}$\@.
(a) The excluded regions in the $(\mixstop, \, \mstop)$ plane
where the mass difference between the $\stopm$ and the $\snu_{\tau}$ 
is greater than 8 or 10~GeV\@.
The dash-dotted straight line shows the kinematic limit 
for this decay.
(b) The excluded regions in the $(\mstop, \, m_{\snu})$ plane,
for a mixing angle of the $\stopm$ assumed to be
0.0 and 0.98 rad.
The solid lines show the actual limits, and 
the thin lines show the limits calculated only with the
expected number of background events. 
The dash-dotted horizontal line shows the limit on $m_{\snu}$
obtained at LEP1~\cite{snulimit,PDG},
and the dash-dotted diagonal line shows the kinematic limit 
for the $\stopm \ra \bq \tau \snu$ decay. 
}
\label{fig:result3}
\end{figure}
%-----------------------------------------------------------------------
\newpage 
\begin{figure}[htb]
\begin{center}\mbox{
\epsfig{file=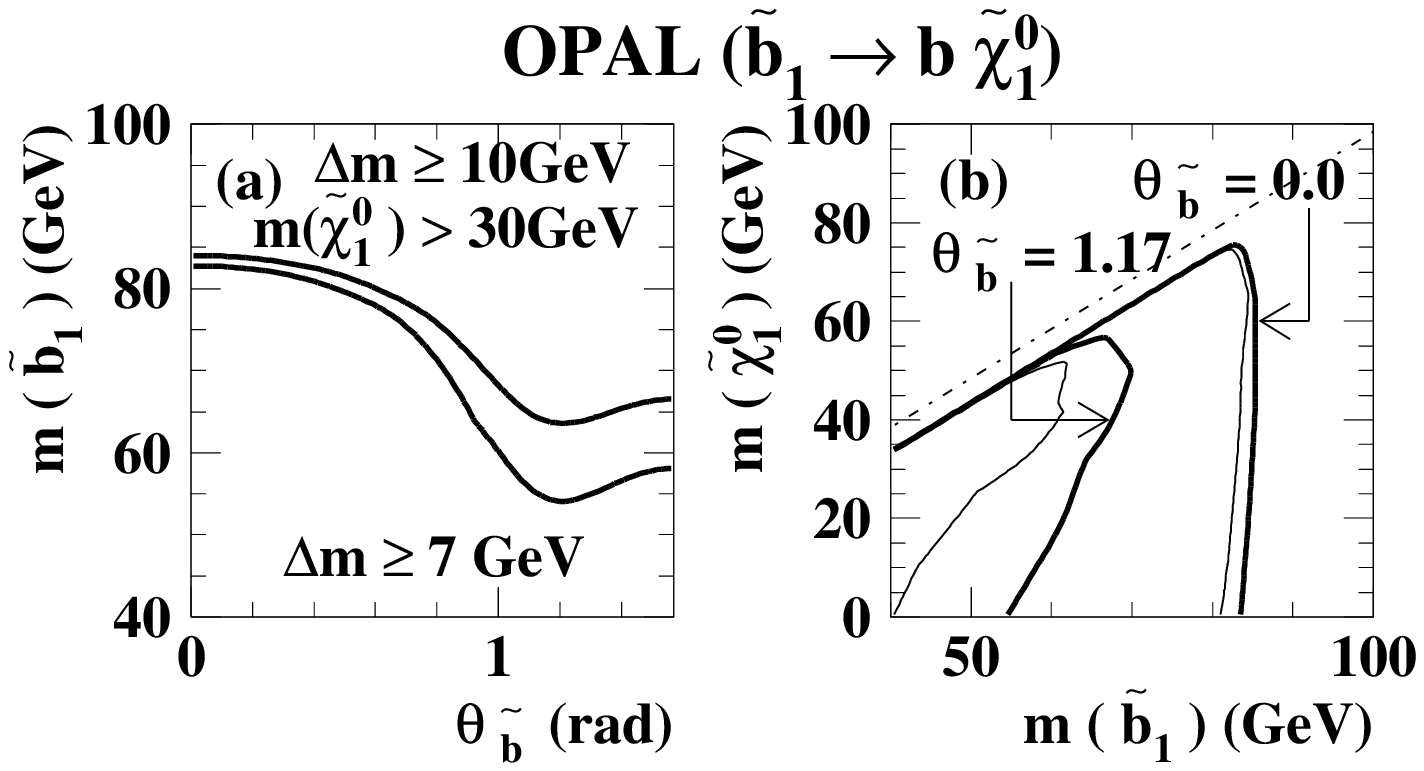,width=16.0cm}
}\end{center} 
\caption[]
{
The 95\% C.L. excluded regions assuming 
that the $\sbotm$ decays into $\bq \neutralino$.\\
(a) The excluded region in the $(\mixsbot, \, m_{\sbotm})$ plane
for a mass difference, $ \Delta m \, ( = m_{\sbotm} - \mchi)$, 
$\Delta m \, \geq$ 10~GeV and $\neutralino$ is heavier than 30~GeV.
The excluded region for $\Delta m \, \geq$ 7~GeV is also shown.\\ 
(b) The excluded regions in the $(m_{\sbotm}, \, \mchi)$ plane,
for a mixing angle of the $\sbotm$ assumed to be 0.0 and 1.17 rad.
The solid lines show the actual limits, and 
the thin lines show the limits calculated only with the
expected number of background events. 
}
\label{fig:result4}
\end{figure}
%-----------------------------------------------------------------------

%%%%%%%%%%%%%%%%%%%%%%%%%%%%%%%%%%%%%%%%%%%%%%%%%%%%%%%%%%%%%%%%%%%%%%%%
\end{document}